\documentstyle[12pt]{article}
\newfont{\bg}{cmr10 scaled\magstep4}
\newcommand{\bigzerol}{\smash{\hbox{\bg 0}}}
\newcommand{\bigzerou}{%
   \smash{\lower1.7ex\hbox{\bg 0}}}
\begin{document}
\title{Mirror Symmetry \\
and \\
An Exact Calculation of N-2 Point Correlation Function\\
on Calabi-Yau Manifold embedded in \(CP^{N-1}\)\\
}
\author{Masao Jinzenji and Masaru Nagura\thanks{A fellow of the Japan Society
for the Promotion of Science for Japanese Junior Scientists } \\
\\
{\it Department of Phisics, The University of Tokyo}\\
{\it  Bunkyo-ku, Tokyo 113, Japan}}
\maketitle

\begin{abstract}

We consider an (N-2)-dimensional Calabi-Yau manifold which is
defined as the zero locus of the polynomial of degree $N$ (of Fermat type)
in {\boldmath$C$}$P^{N-1}$ and its mirror manifold.
 We introduce  the (N-2)-point
correlation function ( generalized Yukawa coupling ) and evaluate it
both by solving the Picard-Fuchs equation for period integrals in the mirror
manifold and also by
explicitly calculating the contribution of holomorphic maps of degree 1
to the Yukawa coupling 
in the Calabi-Yau manifold using the method of Algebraic geometry.
In enumerating the holomorphic curves in the general dimensional
Calabi-Yau manifolds , we extend the method of counting rational curves on
the Calabi-Yau 3-fold using
the Shubert calculus on $Gr(2,N)$.
%
The agreement of the two calculations for the (N-2)-point function
establishes ``{\it the mirror symmetry at the correlation function level} "
 in general dimensional case.
\end{abstract}
\section{Introduction}

In a previous paper \cite{nag} one of the present authers discussed
the mirror symmetry in 2 and 1-dimensional cases ,
i.e. K3 and complex torus . In this paper we generalize the method
of \cite{nag} to the case of arbitrary dimensions.

Specifically we consider an (N-2)-dimensional hypersurface W defined by
\begin{equation}
P=X_1^N + X_2^N + \cdots + X_N^N - N\psi X_1 X_2 \cdots X_N = 0
\end{equation}
in $CP^{N-1}$ which is an (N-2)-dimensional Calabi-Yau manifold
( $1_{st}$ chern index vanishes ). The mirror partner M of this Calabi-Yau
manifold W is constructed by first taking the quatient $W/Z_N^{\otimes N-2} $
and then by blowing up its singulalities . In the following
we compute an (N-2)-point correlation function which is an analogue of the
Yukawa coupling in the both these models.

In the case of the calculation on the mirror manifold M ( B-model ) ,
we use Picard -Fuchs equations for the period integrals.

In the case of the calculation on the original manifold W ( A-model ),
we use a method of Algebraic geometry for the enumeration of
holomorphic curves in projective hypersurface .
We show that the results of the calculations agree and
provide strong evidence for the mirror symmetry in the case of
general dimensions.

We note that the analysis of the Picard-Fuchs equations in general
dimensional Calabi-Yau manifold was given by Green , Morrison and Plesser
\cite{h.d.m.} which this was in preparation. They did not derive the Yukawa
coupling from the A-model and did not establish the agreement of A and B model
calculations .

The aim of the first half of this paper is to calculate the $(N-2)$-point
Yukawa coupling on the B-model exactly. Roughly speaking, the $(N-2)$-point
Yukawa coupling is that
$$
<(X_1 X_2 \cdots X_N)^{N-2}>_{B-model}.
$$
This is a straightforward extension of the Yukawa coupling in
$3$-dimensional case, so we call it the $(N-2)$-point Yukawa coupling.
(Unfortunately, however, it is seems to be hard to give a special mening
such as `` Yukawa Coupling ".) We can formulate the Yukawa coupling
in terms of $\Omega$,
 which is the holomorphic $(N-2,0)$-form on M, and its derivative with
respect to the moduli variable $\psi$.
$$
<(X_1 X_2 \cdots X_N)^{N-2}>_{B-model} =
\int_M \Omega \wedge \frac{\partial^{N-2}}{\partial \psi^{N-2}} \Omega
\;.
$$
Therefore we can calculate the $(N-2)$-point Yukawa coupling immediately
if we obtain $\Omega$ as a function of $\Omega$. We will devote the
first half of this paper to how to calculate $\Omega$ as a function
( or  more exactly as a section ) on the complex moduli space of M.

After obtaining an explicit result of the $(N-2)$-point Yukawa coupling of
the B-model on M, we transform it to that of the A-model on W. All we
have to do in this process is the following;
\newline 1) to convert the complex moduli variable $\psi$ to the k{\" a}hler
moduli variable $t$ which is given by the mirror map,
\newline 2) to fix the gauge condition of $\Omega$ which is a section of
a line bundle on the complex moduli space $\cal M$;
$$
<->_{B-model}\to <->_{A-model}=
{\frac{1}{\Omega_0}}^2<->_{B-model}({\frac{d\psi}{dt}})^{N-2}.
$$

 Next, we interpret the $(N-2)$-point Yukawa coupling from the A-model point
 of view.

 In the A-model, correlation function is calculated as
 intersection number on moduli space of holomorphic maps ( instanton ) from
$CP^{1}$ to the target manifold. We use the fact that moduli space
of degree $k-$instantons can be
approximated as algebraic submanifold of $CP^{(k+1)N+1}$ and show the
principle part of $n^{N}_{k}$ ,which is the $k$-th coefficients of expansion
 of $(N-2)$ point Yukawa coupling,can be derived as the number of solutions
 of a certain set of algebraic equations.

 Correction terms to $n^{N}_{k}$ appear from the degeneration of these
equations. It occurs in two cases.\newline
 case 1) $\phi \in CP^{(k+1)N-1}$ becomes a  holomorphic map of a lower degree
because of projective equivalence.
\newline
 Case 2) $\phi \in CP^{(k+1)N-1}$ is a  multiple cover map.

 Case 2) occurs only if $k\geq 2$. In $k=1$ case, we can remove degeneration
 of case 1) by introducing Grassmann manifold $Gr(2,N)$ instead of $CP^{2N+1}$.
 We construct an exact moduli space of degree 1 with the aid of the Schubert
calculus and  by combining with the algebraic equations, we
explicitly derive $n^{N}_{1}$. $n^{N}_{1}$ exactly agrees with that
calculated from the Picard-Fuchs equation and the mirror map.

 This coincidence of the two calculation ``{\it confirm  the mirror symmetry
at the correlation level in general dimension }" , especially for the case
when
instantons (holomorphic maps) have non trivial moduli degree of freedom.

\section{The B-model}

\subsection{some reviews on the definition of the B-model}
The B-model is obtainned by {\it twisting} a N=2 non-linear
sigma model defined on a Calabi-Yau space \cite{w1}. N=2 non-linear
sigma model is defined as follows.

let M be a n-dimensinal Calabi-Yau manifold and $\phi^i$ be a holomorphic
coordinate on M ($i=1,\cdots n$)(and $\phi^i$ be a anti-holomorphic
coordinate ), $\Sigma$ be a Rieman surface, which, in this paper,
is rescricted to genus zero, and $z$ be a holomorphic coordinate on $\Sigma$.

The Lagrangian is
\begin{equation}
L=2t\int_\Sigma d^2 z (g_{IJ}\partial_z \psi^I \partial_{\bar{z}}\phi^J
+ i\psi_-^{\bar{i}} D_z \psi_-^i g_{{\bar{i}}i} + i\psi_+^{\bar{i}}D_{\bar{z}}
\psi_+^i g{{\bar{i}}i}+ R_{i{\bar{i}}j{\bar{j}}}\psi_+^i \psi_+^{\bar{i}}
\psi_-^j \psi_-^{\bar{j}})
\label{sigmalag}
\end{equation}
where $\phi^i(z)$ is a holomorphic function from $\Sigma$ to M (these are the
main dynamical variables in this model) and $\psi_+^i$ is a section of
$K^{1/2}\otimes \Phi^*(T^{(1,0)}M)$. K is a canonical bundle on $\Sigma$ and
$K^{1/2}$ is its square root. $\Phi^*(T^{(1,0)})$ is a pull-back of
$T^{(1,0)}$ by the map $\phi^i$. Similarly, $\psi_+^i$ is a section of
$K^{1/2}\otimes \Phi^*(T^{(0,1)}M)$, and $\psi_-^i$, $\psi_-^i$ are sections
of $\bar{K}^{1/2}\otimes \Phi^*(T^{(1,0)}M) $, $\bar{K}^{1/2}\otimes \Phi^*
(T^{(0,1)}M)$ respectively.(Of cource, $\bar{K}$ is a anti-canonical bundle on
$\Sigma$.)

This Lagrangian posseses N=2 supersymmetry.In terms of fermionic
parameter $\alpha_-, \tilde{\alpha}_-$, which take values on holomorphic
section of $K^{-1/2}$, and $\alpha_+, \tilde{\alpha}_+$, which take values
on holomorphic section $K^{1/2}$, the supertranslation laws are given
\begin{eqnarray}
\delta \phi^i & = & i\alpha_- \psi^i_+ + i \alpha_+ \psi^i_- \nonumber \\
\delta \phi^{\bar{i}} &=& i\tilde{\alpha}_- \psi_+^{\bar{i}}+
i\tilde{\alpha}_+ \psi_-^{\bar{i}} \nonumber \\
\delta \psi_+^i &=& - \tilde{\alpha}_- \partial_z \phi^i -i \alpha_+
\psi_-^{\bar{j}}\Gamma^i_{{\bar{j}}m}\psi_+^m \nonumber \\
\delta \psi_+^{\bar{i}} &=& - \alpha_- \partial_{\bar{z}} \phi^{\bar{i}} -i
\tilde{\alpha}_+ \psi_-^{\bar{j}} \Gamma^{\bar{i}}_{{\bar{j}}{\bar{m}}}
\psi_+^{\bar{m}} \nonumber \\
\delta \psi_-^i &=& - \tilde{\alpha}_+ \partial_z \phi^i -i \alpha_-
\psi_+^{\bar{j}}
\Gamma^i_{{\bar{j}}m}\psi_-^m \nonumber \\
\delta \psi_-^{\bar{i}} &=& - \alpha_+ \partial_{\bar{z}} \phi^{bar{i}} -i
\tilde{\alpha}_- \psi_+^{\bar{j}} \Gamma^{\bar{i}}_{{\bar{j}}{\bar{m}}}
\psi_+^{\bar{m}}
\label{strns}
\end{eqnarray}
B-model is obtained by twisting the above Lagrangian as follows;
\newline
1) Instead of taking $\psi^i_+$ and $\psi^{\bar{i}}_+$ to be sections
$K^{1/2}\otimes \Phi^*(T^{(1,0)}M)$ and $K^{1/2}\otimes \Phi^*(T^{(0,1)}M)$,
we take them to be sections of $K\otimes \Phi^*(T^{(1,0)}M)$ and $\Phi^*(
T^{(0,1)}M)$.
\newline
2)As to $\phi_+^i$ and $\phi_+^{\bar{i}}$, similarly, we redefine them to be
sections of $\bar{K}\otimes \Psi^*(T^{(1,0)}M)$ and $\bar{K}\otimes \Psi^*
(T^{(0,1)}M)$.

Here for convenience we redefine the variables;
\begin{eqnarray*}
   \eta^{\bar{i}} & = & \psi^{\bar{i}}_+ + \psi^{\bar{i}}_- \\
   \theta_i       & = & g_{i\bar{i}} ( \psi^{\bar{i}}_+ - \psi^{\bar{i}}_- ) \\
   \rho_z^i       & = & \psi^i_+ \\
   \rho_{\bar{z}}^i & = & \psi^i_-
\end{eqnarray*}

The  B-model Lagrangian is
\begin{equation}
L = t \int_\Sigma d^2 z ( g_{IJ} \partial_z \phi^I \partial_{\bar{z}}
\phi^J  + i \eta^{\bar{i}} (D_z \rho_{\bar{z}}^i + D_{\bar{z}} \rho_z^i )
g_{i{\bar{i}}} + i \theta_i(D_{\bar{z}} \rho_z^i - D_z \rho_{\bar{z}}^i)
+ R_{i{\bar{i}}j{\bar{j}}} \rho_z^i \rho_{\bar{z}}^j \eta^{\bar{i}} \theta_k
g^{k{\bar{j}}})
\label{lag-b}
\end{equation}
Since the canonical bunle $K$ (or $\bar{K}$) is trivial, the twisting does
nothing at all at least locally. Therefore the transformation law (\ref{strns})
should be  still valid. But to keep up with the change of the bundl on which
$\psi_+^i$, $\psi_-^i$ etc. take their value, we also  have to change the
bundles of $\alpha_+$, $\alpha_-$ etc. By the B-twisting the infinitesmal
parameter $\alpha_-$,$\tilde{\alpha}_-$, $\alpha_+$, and $\tilde{\alpha}_+$
turn followingly;
$$
\begin{array}{ccc}
\alpha_{-},\tilde{\alpha_{-}}:\;sections\;of\;K^{-1/2}
& \to &
\left\{
\begin{array}{c}
\alpha_{-}:\;section\;of\;K^{-1}\\
\tilde{\alpha_{-}}:\;scalar\;on\;\Sigma
\end{array}
\right.
\end{array}
$$
$$
\begin{array}{ccc}
\alpha_{+},\tilde{\alpha_{+}}:\;sections\;of\;\bar{K^{+1/2}}
& \to &
\left\{
\begin{array}{c}
\alpha_{+}:\;section\;of\;\bar{K^{-1}}\\
\tilde{\alpha_{+}}:\;scalar\;on\;\Sigma
\end{array}
\right.
\end{array}
$$
According to Eguchi-Yang \cite{ey}, reinterpretating the above scalar
transformations by $\tilde{\alpha}_-$, $\tilde{\alpha}_+$ as BRST
transformation, we can obtain a topological field theory.

That is , the BRST transformation is obtained from (\ref{strns}) by setting
$\alpha_- = \alpha_+ = 0$ and setting $\tilde{\alpha}_- = \tilde{\alpha}_+
= \alpha = constant$. The topological transformations are
\begin{eqnarray}
\delta\phi^i &=& 0 \nonumber \\
\delta\phi^{\bar{i}} &=& i\alpha \eta^{\bar{i}} \nonumber \\
\delta\eta^{\bar{i}} &=& \delta \theta_i = 0 \nonumber \\
\delta\rho^i&=& -\alpha d\phi^i
\label{stb}
\end{eqnarray}
Also we can introduce the BRST operator $Q$ which generate topological
transformation such that $\delta W =-i\{ Q,V \}$ for any field $V$.
$Q$ satisfies the condition $Q^2 =0$ mduloe the equation of motion.

In terms of this BRST operator $Q$, we can rewirte the Lagrangian
(\ref{lag-b}) as;
$$L =it \int \{ Q,R\} + tW \;,$$
where
$$R = g_{i{\bar{j}}}(\rho_z^i \partial_{\bar{z}} \phi^{\bar{j}}+
\rho_{\bar{z}}^i \partial_z \phi^{\bar{j}})\;,$$
and
$$W = \int_{\Sigma}( -\theta_i D \rho^i -i/2 R_{i{\bar{i}}j{\bar{j}}}
\rho^i \wedge \rho^j \eta^{\bar{i}} \theta_k g^{k{\bar{i}}})\;,
$$
here $D$ is the exterior derivative on $\Sigma$ and extended to act
on forms with values in $\Phi^*(T^{1,0}M)$ by using the pull-back of the
Levi-Civita connection on M.

\subsection{The Observables}

While the BRST-invariant observables of A-model form De-Rahm(Dolbeaut)
cohomology on W, those of B-model are given by $\bar{\partial}$-cohomology
of \newline
$A^p (M,\wedge^q T^{1,0}M)$ : $A^p (M,\wedge^q T^{1,0}M)$ is the set of
(0,p)forms on M which take values in $\wedge^q T^{1,0}M$. (Here $
\wedge^q T^{1,0}M$ means the q-th exterior power of holomorphic tangent
bundle  on M $T^{1,0}M$.) Such an object can be written;
\begin{equation}
A^p (M,\wedge^q T^{1,0}M \ni V =V_{{\bar{i}}_1  {\bar{i}}_2 \cdots
{\bar{i}}_p}^{j_1 j_2 \cdots j_q }dz^{{\bar{i}}_1}dz^{{\bar{i}}_2}
\cdots dz^{{\bar{i}}_p} \partial_{j_1} \partial_{j_2} \cdots
\partial_{j_q}
\end{equation}
Of course, the sheaf cohomology group $H^p(M,\wedge^q T^{1,0}M)$ is defined by
the quatient space (module) of $Z^p(M,\wedge^q T^{1,0}M)$ which is the solution
space of $\bar{\partial}V=0$ moduloe $B^p(M,\wedge^q T^{1,0}M)$ which is the
set of $S$'s such that $\bar{\partial}S=0$ for all $S\in A^{p-1}(M,\wedge^q
T^{1,0}M)$.

Given any point $P\in\Sigma$, we can give the correspondance of every elements
$V$ of $A^{p-1}(M,\wedge^q T^{1,0}M)$ to the quantum field theory operator
${\cal O}_V (P)$;
\begin{equation}
{\cal O}_V (P) = V =V_{{\bar{i}}_1  {\bar{i}}_2 \cdots
{\bar{i}}_p}^{j_1 j_2 \cdots j_q }(P)
\eta^{\bar{i}_1}\eta^{\bar{i}_2} \cdots \eta^{\bar{i}_p}
\psi_{\bar{j}_1} \psi_{\bar{j}_2} \cdots \psi_{\bar{j}_q},
\end{equation}
and we can find that
$$ \{Q,{\cal O}_V \} = -{\cal O}_{\bar{\partial}V}. $$

Therfore ${\cal O}_V$ is BRST-invariant if and if only $\bar{\partial}V=0$,
and
${\cal O}_V$ is BRST-exact if and if only $V=\bar{\partial}S$ for some $S$.

this correspodance gives a natural map from $\oplus_{p,q=0}^{p,q=n}H^p
(M,\wedge^q T^{1,0}M)$ to the BRST cohomology or the observables of B-model.
This map, in fact, is isomorphic.

\subsection{(N-2)-point Yukawa coupling}
We describe the observable which corresponds to the representive
$V_a \in H^{P_a}(M,\wedge^{Q_a} T^{1,0}M)$ as ${\cal O}_{V_a}$.
Our concern is the formulation of the correlation function
\begin{equation}
<\prod_a {\cal O}_{V_a}(P_a)>,
\label{corf}
\end{equation}
where $P_a$'s are arbitrary points on $\Sigma$.

The operator ${\cal O}_{V_a}$ has a left-moving ghost number $P_a$
and a right-moving one $Q_a$. In order for (\ref{corf}) not to vanish,
the conservation law between the left(or right)-movin ghost number and
back ground ghost charge requests that
\begin{equation}
\sum_a P_a = \sum_a Q_a = n(1-g),
\end{equation}
where $g$ means the genus of $\Sigma$ and $n$ is the complex dimension of
the target space M.

In this paper we especially concern the correlation function of the type
$$<\prod_{i=1}^n {\cal O}_B (P_i)>,$$
where $B\in H^1(M,T^{1,0}M)$.
We would like to call this kind of corretion function {\it n-point Yukawa
coupling} from now on.

\subsection{Path-Integral evaluation -- Reduction to Constant Maps}

According to Vafa \cite{vafa} and Witten \cite{w1} it is revealed
that the path-integral of B-model can
be reduced to the integral on the space of the constant maps from the
worldsheet $\Sigma$ to its target M, i.e.  the integral on the target space M
itself. The {\it physical proof} is the following .

Let $\Upsilon$ be some function space on which we wish to
path-integrate. Consider the theory we are dealing with (of cource, in
this case B-model) has a group symmetry $G$. Suppose that $G$ acts freely
on $\Upsilon$. Then there is a fibration $\Upsilon \to \Upsilon / G$, and
we can reduce the path-integral over $\Upsilon$ to $\Upsilon / G$.
If we consider only $G$ invariant observables $\cal O$, we have the
formula,
\begin{equation}
\int_\Upsilon {\cal D}\phi e^{-S} {\cal O} = (volG) \int_{\Upsilon / G}
{\cal D}\phi e^{-S} {\cal O},
\label{rdc}
\end{equation}
where $volG$ is the volume of the group $G$.

We can apply this formula to the case in which $G$ is a supergroup
gererated by the BRST charge $Q$. but this case is rather strange,
since for any fermionic variables $\theta$
$$\int d\theta \cdot 1 = 0, $$
the volume of any supersymmetry $G$ becomes zero.
Do we have to conclude any correlation function of BRST invariant
operators are all zero ?

In general, the group $G$ does not act {\it freely}.
In almost all cases
there are some fixed point sets $\Upsilon_0$. The nonzero contribution to the
correlation function comes from only $\Upsilon_0 $ .
Since on $\Upsilon - \Upsilon_0$
$G$ acts {\it freely} then we can apply (\ref{rdc}) there.

In the B-model $Q$-invariant points must satisfy from (\ref{stb}) that
$$ \delta \rho^i = - \alpha d\phi^i = 0 ,$$
therefore $$d\phi=0.$$
This means that the maps $\Phi :\Sigma \to M $ should be constant maps on
M. Thus we have succeeded in reducing the path-integral over $\Upsilon$
to the integral
over the space of constant map on M , that is , the integral on target space M
itself.

Next we want to consider how to calculate n-point Yukawa couping
\newline
$<{\mathop{\prod}_{i=1}^d}{\cal O}_B(P_i)>$ concretely.
\begin{eqnarray*}
{\cal O}_B(z_k) & = & b_{\bar{i}}^j \eta^{\bar{i}} \theta_j \\
{\mathop{\prod}_{k=1}^n}{\cal O}_B(z_k) & = & b_{\bar{i_1}}^{j_1}
\cdot (\eta^{\bar{i_1}}\theta_{j_1}) \\
    & \equiv & \theta_Y .
\end{eqnarray*}
Corresponding to the map $\otimes^n \theta_B \mapsto \theta_Y$,
we can construct the map:
\begin{eqnarray*}
B^{\otimes^n} & \mapsto & Y \\
\\
\otimes^n H^1(M,T^{1,0}M) & \to & H^n (M,\wedge^n T^{1,0}M),
\end{eqnarray*}
It is apparent that this map is merely a classical wedge product.

Now in order to carray out the integral over Calabi-Yau manifold M,
we need to transform the element of $H^n(M,\wedge^n T^{1,0}M)$ to
the element of $H^n(M,\Omega^n M)$ (here $\Omega^n M$ means the sheaf of
(n,0)form.)

We can realize the requested transformation by operating the square of
the holomorphic (n,0)-form on M. According to the general theory of Calabi-Yau
manifold, the holomorphic (n,0)-form on n-dimensional M exist uniquely up to
constant. So we don't have to worry about how to select the holomorphic forms.
Therfore we can formulate n-point Yukawa coupling up to constant by the
integral on M itself,
\begin{equation}
<{\mathop{\prod}_{i=1}^n}{\cal O}_B(z_i)> =
\int_M \Omega \wedge b^{i_1} \wedge b^{i_2} \wedge \cdots
\wedge b^{i_n} \Omega_{{i_1} {i_2} \cdots {i_n}},
\end{equation}
where
\begin{eqnarray*}
b^{i_k} & = & b^{i_k}_{\bar{j}}dz^{\bar{j}} \\
\Omega  & = & \Omega_{{j_1} {j_2} \cdots {j_n}}dz^{i_1}dz^{i_2}
\cdots dz^{i_n} \in H^{n,0}(M).
\end{eqnarray*}
Note that this formula is defined only up to a constant.

\subsection{Kodaira-Spencer equation}

It is known that $b^i \in H^1_{\bar{\partial}}(M,T^{1,0}M), i=1 \cdots
dim H^1_{\bar{\partial}}(M,T^{1,0}M)$ form a basis of the tangent space
to the moduli space
of the complex structure of M ( we will denote it  ${\cal M}$ );
$$ T{\cal M}|_M = H^1_{\bar{\partial}}(M,T^{1,0}M).$$
Kodaira and Spencer \cite{ks1} \cite{ks2} showed that the complex-structure
moduli
space $\cal M$ itself is also a complex manifold. Let $z^{\alpha} (\alpha =1
\cdots dim{\cal M}=dimH^1(M,TM))$ be a holomorphic coordinate on $\cal M$
(In this paper we will deal only with the case $dim{\cal M}=1$.)

The deformation equation of Kodaira and Spencer is the following;
\begin{equation}
\frac{\partial \Omega}{\partial z^{\alpha}} =
k_{\alpha} \Omega + \chi_{\alpha},
\label{ks-eq}
\end{equation}
where $k_{\alpha}$ depends only on $z_{\alpha}$, but not on the coordinate
of M, and $\chi_{\alpha} \in H^{n-1,1}(M)$. Thas is , this means a
decomposition:
$$\frac {\partial \Omega}{\partial z^{\alpha}} \in H^{n,0}\oplus H^{n-1,1}
. $$
Let us  guess what happens here. We can describe the holomorphic (n,0)-form
$\Omega$ in terms of a holomorphic local coordinate $x^{\mu}$ of M as
\begin{equation}
\Omega = \frac{1}{n!}h(x)\epsilon_{{\mu}{\nu}\cdots {\rho}}dx^{\mu}dx^{\nu}
\cdots dx^{\rho}.
\label{Omega}
\end{equation}
Noting that the holomorphic local coordinate $x^{\mu}$ depends on the
complex structure $z^{\alpha}$, derivate the both sides of (\ref{Omega})
with $z^{\alpha}$. Then
\begin{equation}
\frac{\partial \Omega}{\partial z^{\alpha}} =
\frac{1}{n!}\frac{\partial h(x)}{\partial z^{\alpha}}
\epsilon_{{\mu}{\nu}\cdots {\rho}}dx^{\mu}dx^{\nu}\cdots dx^{\rho}
+
\frac{1}{(n-1)!}h(x)
\epsilon_{{\mu}{\nu}\cdots {\rho}}\frac{\partial dx^{\mu}}{\partial z^{\alpha}}
dx^{\nu}\cdots dx^{\rho}.
\end{equation}
The first term is apparently a pure (n,0)-form.
But the second term is a direct linear combination of (n,0)-form and
(n-1,1)-form, since, especially we should note the term
$\frac{\partial dx^{\mu}}{\partial z^{\alpha}}$ ,
this term consists of (1,0)-form part and (0,1)-form part.

thus,
\begin{eqnarray}
\chi_{\alpha} & = & \left. \frac{\partial \Omega}{\partial
z^{\alpha}}
\right|_{(n-1,1)form \; part} \nonumber \\
 & = & \frac{1}{(n-1)!} h(x)\epsilon_{{\mu}{\nu}\cdots {\rho}}
\left.\frac{\partial dx^{\mu}}{\partial z^{\alpha}}
\wedge dx^{\nu}\wedge \cdots \wedge dx^{\rho}
\right|_{(n-1,1)form \; part} \nonumber \\
 & = &  \frac{1}{(n-1)!} h(x)\epsilon_{{\mu}{\nu}\cdots {\rho}}
\left. \frac{\partial dx^{\mu}}{\partial z^{\alpha}}
\right|_{(0,1)form \; part}
\wedge dx^{\nu}\wedge \cdots \wedge dx^{\rho}\
\label{modifiedks-eq1}
\end{eqnarray}
It is also known that $H^{n-1,1}(M)$ and $H^1_{\bar{\partial}}(M,TM)$ are
ismorphic each other by the help of the holomorphic (n,0)-form $\Omega$.
There is the map from $H^1_{\bar{\partial}}(M,TM)$ to $H^{n-1,1}(M)$;
\begin{eqnarray*}
H^1_{\bar{\partial}}(M,TM) & \to & H^{n-1,1}(M) \\
\\
\chi_{\alpha}^{\mu}=\chi_{{\alpha},\; {\bar{\nu}}}^{\mu}dx^{\bar{\nu}}&\mapsto&
\chi_{\alpha}=\chi_{\alpha \; {\bar{\nu}}}^{\mu}
\Omega_{{\mu} {\rho} \cdots {\sigma}}
dx^{\bar{\nu}}dx^{\rho}\cdots dx^{\sigma}.
\end{eqnarray*}
We can inserse this map;
\begin{eqnarray*}
 H^{n-1,1}(M)& \to & H^1_{\bar{\partial}}(M,TM) \\
\\
\chi_{\alpha}= & \chi_{\alpha,\; {\bar{\nu} {\mu} {\rho} \cdots
{\sigma}}} dx^{\bar{\nu}}dx^{\rho}\cdots dx^{\sigma}
 \mapsto&
\chi_{\alpha}^{\mu}=\frac{1}{2||\Omega||^2}\bar{\Omega}^{{\mu}{\rho}\cdots
{\sigma}}\chi_{\alpha,\; {\bar{\nu} {\rho} \cdots
{\sigma}}} dx^{\bar{\nu}}
\end{eqnarray*}
The original Kodaira-Spencer equation (\ref{ks-eq}) can be rewritten
in terms of the element of $H^1_{\bar{\partial}}(M,TM)$,
\begin{equation}
\frac{\partial \Omega}{\partial z^{\alpha}}=
k_{\alpha}\Omega + \chi_{\alpha}^{\mu}\Omega_{{\mu} \cdot}\;\;,
\end{equation}
where $\chi_{\alpha}^{\mu} \in H^1_{\bar{\partial}}(M,TM)$,
and $\Omega_{{\mu} \cdot}$ means that
\begin{equation}
\Omega_{{\mu} \cdot}\equiv \Omega_{{\mu}{\nu} \cdots {\rho} }
dx^{\nu}\wedge  \cdots \wedge dx^{\rho};(n-1)form .
\label{modifiedks-eq2}
\end{equation}

{}From (\ref{modifiedks-eq1}) and (\ref{modifiedks-eq2})
we can infer that
\begin{equation}
\chi_{\alpha}^{\mu}=
\left. \frac{\partial dx^{\mu}}{\partial z^{\alpha}}
\right|_{(0,1)form \; part}.
\end{equation}
Therfore from all the fact above we can  immediately derive
that
\begin{eqnarray}
"n-point \; & Yukawa \; coulping " \nonumber \\
 \equiv  & \int_{\Sigma} \Omega \wedge b_{\alpha}^{i_1} \wedge
b_{\beta}^{i_2} \wedge \cdots \wedge b_{\gamma}^{i_n}
\Omega_{{i_1}{i_2} \cdots {i_n}}\\
 = & \int_{\Sigma} \Omega \wedge
\underbrace{
\frac{\partial^n \Omega}
{\partial z^{\alpha}\partial z^{\beta}
\cdots \partial z^{\gamma}}
}_{\mbox{$n$ times}}.
\label{Y.C.}
\end{eqnarray}
It is because that obviously from (\ref{modifiedks-eq1})
\begin{equation}
\left.
\underbrace{
\frac{\partial^n \Omega}
{\partial z^{\alpha}\partial z^{\beta}
\cdots \partial z^{\rho}}
}_{\mbox{$m$ times}}
\right|_{(n-m,m)form \; part}
=
\chi_{\alpha}^{i_1}\wedge \chi_{\beta}^{i_2}\wedge
\cdots \wedge \chi_{\beta}^{i_m}
\Omega_{{i_1}{i_2}\cdots {i_m}. },
\end{equation}
where $\Omega_{{i_1}{i_2}\cdots {i_m}. }\stackrel{\rm def}{=}
\Omega_{{i_1}{i_2}\cdots {i_m}{j_1}{j_2}\cdots {j_{n-m}}}
dx^{j_1}dx^{j_2}\cdots dx^{j_{n-m}}$.
we should also note that after integrating over M only $(0,n)$form
part of $\underbrace{
\frac{\partial^n \Omega}
{\partial z^{\alpha}\partial z^{\beta}
\cdots \partial z^{\gamma}}
}_{\mbox{$n$ times}}$
remains non-zero.

Thus all we have to do for the calculation of the n-point Yukawa
coupling is to calculatr the holomorphic (n,0)form $\Omega$
on M as a function on the complex-structure moduli space
(, roughly speaking, in our model as a function of $\psi$).

\subsection{The periods}
{}From the arguement of the previous section we have concluded that we can
calculate n-point Yukawa coupling of n-dimensional B-model
if we can get the holomorphic $(n,0)$-form on the target Calabi-Yau
space as a function on the moduli space (or, more precisely, as a
section of the line bundle on the moduli space).

Our main aim in this subsection is to compute concretely the holomorphic
$(n,0)$-form as a function $\psi $.

We want to concern the following Calabi-Yau manifold M which is defined
in $CP^{N-1}$ as a zero locus of the homogeneous polynomials of degree
$N$;
\begin{equation}
W=X^N_1+X^N_2+ \cdots + X^N_N - N \psi X^N_1 X^N_2\cdots X^N_N .
\end{equation}
(To be more pricise we have to divide the zero locus by
$Z^{\otimes^{N-2}}_N$ and blow up the singularities. However this
modification has the effect only multiplying the preods by the over all
factor ${(\frac{1}{N}})^{N-2}$.)

\subsubsection{General theory of the Picard-Fuchs equation}
According to the gerenal theory of Algebraic geometry a trivial
( or nowhere zero)
holomorphic $(n,0)$-form $\Omega$ on a n-dimensionsl manifold N
exists if its first chern index vanishes. That the first chern index vanishes
is the very condition to be Calabi-Yau manifold.
 This holomorphic $(n,0)$-form,
say, $\Omega$ is given in terms of the defining equation $W$ ;
\begin{eqnarray*}
\Omega & = & \int_{\gamma}\frac{1}{W}\omega, \\
& \omega & \equiv  \sum_{A=1}^{N}(-1)^A X^A
dX^1 \wedge \cdots \wedge {\check {dX^A}} \wedge \cdots \wedge dX^N ,
\end{eqnarray*}
where $\gamma$ is a small and one-dimensional cycle winding around the
hypersurface $N=\{(X_1:X_2:\cdots :X_N)\in CP^{N-1}|\;W(X_1,X_2,\cdots ,X_N)
=0\}$.  Thus $\Omega$ can be defined in practice, but it is difficult or almost
imposible to carry out this integral.
The Picard-Fuchs equation is a good means of escape from this difficulity.

Introduce the integral
\begin{equation}
\Omega_k = \int_{\gamma}\frac{P_k (X)}{W}\omega ,\;\; k=0\cdots N-2 ,
\label{p1}
\end{equation}
where $P_k (x)$ is a homogeneus polynomial of degree $k(N-2)$.
This integral represents an element of $\oplus_{q=0}^k \wedge^{(N-2-q,q)}$.
In \cite{l.s.w.} it is showed that the $\Omega_{\alpha}$ represents
a nontrivial cohomology element of
$F^k \equiv \oplus_{q=0}^k H^{(N-2-q,q)}(M;R)$
if and only if $P_k $ is a nontrivial element of the local ring or
the chiral ring $\cal R$ of W. If we take $P_{\alpha}$'s to be a basis for
$\cal R$, the corresponding $\Omega_k$'s form a basis for $H^{N-2}(M;R)$.
(In this paper, we take a subring of $\cal R$, so the corresponding
$\Omega_k$'s form a basis for the corresponding subcohomology ring of
$H^{N-2}(M;R)$.)

The set of periods of differential form $\Omega_k$ is defined as the
integral of $\Omega_k$ over elements of a basis of the integral homology
on M.
$$
\Pi_{\alpha}^{\beta} = \int_{\Gamma_{\beta}}\Omega_{\alpha}
= \int_{\Gamma_{\beta}\times \gamma}\frac{P_{\alpha} (X)}{W}\omega,
$$
where $\Gamma_{\beta}$ is a representives of a homology basis $H_{N-2}(M;Z)$,
$\Gamma_{\beta}\times \gamma$ can be considered as a tube over the cycle
$\Gamma_{\beta}$. The set of periods is a matrix-valued function on the moduli
spce.

Now we fix the homology cycle $\Gamma_{\beta}$ and define a vector
$\omega_{\alpha}\equiv \Pi_{\alpha}^{\beta}$. we call this vector
{\it periods vector}.
We will derive a differential equation for the periods vector.

\subsubsection{The Picard-Fuchs equation for the Fermat type}
Remind that in this paper we are dealing  with the case
$$
W=X_1^N+X_2^N+ \cdots +X_N^N-N\psi X_1^NX_2^N\cdots X_N^N.
$$
The moduli variables in this case is $\psi$.
Let us define the following periods vector $\bar{\omega}\equiv
{(\omega_0,\cdots,\omega_{N-2})}^T $ ;
\begin{equation}
\omega_l = (l-1)!\int_{\Gamma}\frac{{(X_1^NX_2^N\cdots X_N^N)}^l}{W^{l+1}}
\omega ,\;\; l=0 \cdots N-2,
\label{p2}
\end{equation}
here we have picked and fixed some homology cycle of $H_{N-2}(M)$ and
the factor $(l-1)!$ is merely a convention.
Differentiating both sides of (\ref{p2}) with $\psi$ and integrating
the r.h.s. by part, then we can obtain the following formula
( this derivation is somewhat less obvious \cite{l.s.w.} );
\begin{equation}
\left[
(1-\psi^N)\frac{\partial^{N-1}}{\partial \psi^{N-1}}-
\sum_{l=1}^{N-2}b_l^{(N)}\psi^{l+1}\frac{\partial^l}{\partial \psi^l}
\right]
\omega_i = 0 ,\;\;\;
for\;all\;i=0 \cdots N-2.
\label{p4}
\end{equation}
The last differential equation is a regular singular type and
 is called ``{\it Picard-Fuchs equation}". Under the substitution
$\psi^N \to z^{-1}$ and $\omega_i \to z^{1/N}W$,
this equation is transformed into an equation ,
\begin{equation}
\left[
(z\frac{\partial}{\partial z})^{N-1}-
z(z\frac{\partial}{\partial z}+\frac{1}{N})
(z\frac{\partial}{\partial z}+\frac{2}{N})
\cdots
(z\frac{\partial}{\partial z}+\frac{N-1}{N})
\right]
W=0
\label{p3}
\end{equation}
This is a gereralized hypergeomerric differential equation with regular
singular
points at $z=0,\;1$ and $\infty$. Note that $z=0$ is a point of the maximally
unipotent monodromy. That is to say, the index equation around $z=0$ is
$\nu^{N-1}=0$ where $\nu$ is the index for series solution around $z=0$.

\subsubsection{The solutions of The Picard-Fuchs equation}
We will adopt the D.Morrison's recipie \cite{morrison} for the
construction of the mirror map on Calabi-Yau $3$-fold,
also in our $(N-2)$-dimensional case. At first we will calculate the series
solution around $z=0$.

We substitute a series solution $W^0=\sum_{n=0}^{\infty}a_n z^n$  into
(\ref{p3}),and obtain a recursion relation such that
\begin{eqnarray}
a_n & = & \frac{(N(n-1)+1)(N(n-1)+2)\cdots (N(n-1)+N-1)}
{n^{N-1} N^{N-1}}
a_{n-1} \\
(& n & =1,2,\cdots ) \nonumber
\end{eqnarray}
Fixing the first term as $a_0=1$,
we obtain that
\begin{equation}
a_n = \frac{(Nn)!}{(n!)^N N^{Nn}}.
\label{zen}
\end{equation}
Thus the series solution around $z=0$ is
\begin{equation}
W^0 = \sum_{n=0}^{\infty}\frac{(Nn)!}{(n!)^N N^{Nn}}z^n.
\label{w0}
\end{equation}

The Picard-Fuchs equation (\ref{p3}) has (N-1)solutions with singularities
around $z=0$ such as $(\log z)^0,(\log z)^1,\cdots,(\log z)^{N-1}$,
since it is a ordinary differential equation of degree (N-1).
Now we want to obtain all of them.  We introduce the following ansatz;
\begin{equation}
W_x \equiv \sum_{n=0}^{\infty}\frac{\{N(n+x)\}!}{{(n+x)!}^N N^{N(n+x)}}
z^{n+x}.
\label{wx}
\end{equation}
In other words we have shifted all the $n$ in (\ref{w0}) to $n+x$.
Since (\ref{wx}) satisfies (\ref{zen}) for $n\geq 1$, $W_x$ satisfies
that
$$
\left[
(z\frac{\partial}{\partial z})^{N-1}-
z(z\frac{\partial}{\partial z}+\frac{1}{N})
(z\frac{\partial}{\partial z}+\frac{2}{N})
\cdots
(z\frac{\partial}{\partial z}+\frac{N-1}{N})
\right]
W_x
= \frac{(Nx)!}{(x!)^N N^{Nx}}X^{N-1}z^x .
$$
Differentiate both sides of the above equation i-th times with $x$, then set
$x=0$. Noting that the right-hand side becomes zero for $ 0\leq i\leq N-2$,
we find that $\left.
\partial_x^i W_x
\right|_{x=0}$ for $0\leq i\leq N-2$ is a solution of the Picards-Fuchs
equation (\ref{p3}). Further calculation shows that
\begin{equation}
W^i \equiv \left.\partial_x^i W_x\right|_{x=0}=
\sum_{j=0}^{i}{}_iC_j \sum_{n=0}^{\infty}\frac{\partial^j}{\partial x^j}
a_n(0){\frac{z}{N}}^n (\log{\frac{z}{N}})^{i-j},
\end{equation}
where
$$ a_n(x)\equiv \frac{(N(n+x))!}{{(n+x)!}^N}=
\frac{\Gamma (N(n+x)+1)}{{\Gamma (n+x+1)}^N} .$$
That means $\left.\partial_x^i W_x\right|_{x=0}$ has the singularity of
$(\log z)^i$ at $z=0$.

\subsubsection{Construction of the holomorphic $(N-2,0)$-form $\Omega$}
Recall that the Picards-Fuchs equation is a differential equation satisfied
by the periods vector $\omega_{\alpha}$ . The periods vector has been defined
as the following;
\begin{eqnarray*}
\omega_{\alpha} & \equiv & \Pi_{\alpha}^{\beta}=\int_{\Gamma}\Omega_{\alpha}
,\;\;\;\Gamma : fixed \\
 & \Omega_{\alpha} & \in \bigoplus_{q=0}^{\alpha}H^{(N-2-q,q)}(M) \\
 & \Gamma_{\beta} & \in H_{N-2}(M;Z)
\end{eqnarray*}
In other words $\omega_{\alpha}$ is a integral of the element of a basis
$\Omega_{\alpha}$ $(\alpha = 1 \cdots dim H^{N-2}(M))$ for the cohomology
ring $H^{N-2}(M)$ over a fixed homology cycle $\Gamma_{\beta}$.But it is
apparent from the derivation of the Picards-Fuchs equation (Gauss-Manin eq.)
that the Picards-Fuchs equation does not depend on the choice of the
homology cycle $\Gamma_{\beta}$. Thefore the periods matrix
$\Pi_{\beta}^{\alpha}\;\;(\alpha,\beta=1\cdots dim H^{N-2}(M))$ also
satisfies the Picards-fuchs equation.

Thus the following vector such that
\begin{eqnarray}
\hat{\omega}_{\beta}& = & \int_{{\Gamma}_{\beta}}\Omega,\;\;\;
\beta=1 \cdots dim H^{N-2}(M) \\
 & \Omega & \in H^{(N-2,0)}(M),\nonumber
\end{eqnarray}
where ${\Gamma}_{\beta}$'s form a basis for the homology ring $H_{N-2}(M;Z)$,
also must satisfy the Picards-Fuchs equation (\ref{p4}) or (\ref{p3}).
In the previous section we have obtained the complete set of the
solutions of (\ref{p3}),
that is;
$$
W^i =
\left.
\partial_{\alpha}^i W_x
\right|_{x=0} ,
\;\;\; i=0 \cdots N-2 .
$$
There must be a homology cycle $A_i \in H_{N-2}(M)$ such that
\begin{equation}
W^i = \int_{A_i}\Omega.
\end{equation}
Let the cohomology element dual to the homology cycle $A_i$ be
$\alpha^i$, then the holomorphic $(N-2,0)$-form $\Omega$ can be expanded,
in terms of $\alpha^i$, to
\begin{equation}
\Omega =\sum_{i=0}^{N-2} W^i \alpha^i.
\end{equation}
We can infer the intersection number between $A_i$'s by considering the
monodoronomy around $z=0$. Under the transformation $z \to z e^{i2\pi}$,
the $W^i$'s transform in the following way;
\begin{eqnarray}
W^0 &\to    & W^0 \nonumber \\
W^1 &\to    & W^1 + W^0 \nonumber \\
W^2 &\to    & W^2 + 2W^1 +W^0 \nonumber \\
    &\vdots &  \nonumber \\
W^i &\to    & W^i + {}_iC_1 W^1 +{}_iC_2 W^0 +\cdots +W^0 \nonumber \\
    & =     & \sum_{j=0}^i {}_iC_j W^j \\
    &\vdots & \nonumber.
\end{eqnarray}
Therfore corresponding to the above transformation the cohomology basis
$\alpha^i$ changes as
\begin{equation}
\left(
\begin{array}{c}
\alpha^0 \\ \alpha^1 \\ \vdots \\ \alpha^i \\ \vdots \\ \alpha^{N-2}
\end{array}
\right)
\to
\left(
\begin{array}{cccccc}
         1 & 1 & \cdots &    &  \cdots  & 1 \\
           & 1 & \cdots &    &  \cdots  & {}_{N-2}C_1 \\
           &   & \ddots &    &          &  \vdots          \\
           &   &        &  1 & \cdots   & {}_{N-2}C_i\\
           &   &        &    & \ddots   &  \vdots \\
 \bigzerol &   &        &    &          &  1
\end{array}
\right)
\left(
\begin{array}{c}
\alpha^0 \\ \alpha^1 \\ \vdots \\ \alpha^i \\ \vdots \\ \alpha^{N-2}
\end{array}
\right)
=
M
\left(
\begin{array}{c}
\alpha^0 \\ \alpha^1 \\ \vdots \\ \alpha^i \\ \vdots \\ \alpha^{N-2}
\end{array}
\right)
\end{equation}
Since this transformation is a kind of modular transformations,the
intersection number between $\alpha^i$'s must be preserved under the
transformation. The most natural form of such intersection numbers is
\begin{equation}
\left(
\begin{array}{cccccc}
 \bigzerou  &        &                    &        &        & 1 \\
            &        &                    &        & -(N-2) &   \\
            &        &                    & \cdots &        &   \\
            &        & (-1)^i {}_{N-2}C_i &        &        &   \\
            & \cdots &                    &        &        &   \\
 (-1)^{N-2} &        &                    &        &        & \bigzerol
\end{array}
\right)
\equiv
I
\end{equation}
We can verify straightfowardly that
$$
{}^tM I M =I.
$$
Thus $\alpha^i$'s can be renormalized  such a way that
\begin{equation}
\Omega = W^0 \tilde{\alpha}^0 + \sqrt{N-2}W^1 \tilde{\alpha}^1 + \cdots
+ \sqrt{{}_{N-2}C_i}W^i \tilde{\alpha}^i + \cdots + W^{N-2}
\tilde{\alpha}^{N-2},
\label{Omg}
\end{equation}
and
\begin{equation}
\int_M \tilde{\alpha}^i \wedge \tilde{\alpha}^j
= (-1)^i \delta_{N-2,\; i+j} \;,
\label{intersection}
\end{equation}
or
\begin{equation}
  {}^{\#}(\tilde{A}_i \cap \tilde{A}_j) =  (-1)^i \delta_{N-2,\; i+j}\;,
\end{equation}
where, of cource, $\tilde{A}_i$ is a homology cycle dual to $\tilde{\alpha}^i
$. We should remind  here the fact that the differential form of even degree
commute each other, while that of odd degree anticommute .
In case of $N\;=\;even$, where M is even dimensional, $\tilde{\alpha}^i$'s are
forms of degree even, then $\int_M \tilde{\alpha}^i \wedge \tilde{\alpha}^j =
\int_M \tilde{\alpha}^j \wedge \tilde{\alpha}^i$. On the contrary, in case of
$N\;=\; odd$, where M is odd diensional,  $\tilde{\alpha}^i$'s are
forms of degree odd,then $\int_M \tilde{\alpha}^i \wedge \tilde{\alpha}^j =
- \int_M \tilde{\alpha}^j \wedge \tilde{\alpha}^i$.
We can immediately verify that is consistent of the symplectic basis.
\footnote{If we want to make an analogy to the , so called, A- and B-cycle;
$$
{}^{\#}(A_i\cap B_j)=\delta_{ij},\;\;\;
{}^{\#}(A_i\cap A_j)=0 ,\;\;{}^{\#}(B_i\cap B_j)=0,
$$
we only have to divide them  into cases.
\newline Case 1) N : even

Adapt the following assignment of the sign such that
$$
\Omega={\tilde W}_0 \alpha^0+{\tilde W}_1 \alpha^1+
\cdots {\tilde W}_{N/2-2} \alpha^{N/2-2}
+{\tilde W}_{N/2-1} \gamma^{N/2-1}
-{\tilde W}_{N/2} \beta^{N/2}+{\tilde W}_{N/2+1} \beta^{N/2+1}-\cdots
\pm {\tilde W}_{N-1} \beta^{N-1},
$$
where ${\tilde W}_i$ is defined as
$$
{\tilde W}_i=\sqrt{{}_{N-2}C_i}W_i \;,
$$
the sign of the last term is
$$
\left\{
\begin{array}{c}
+ \;\;\; for \;\; the\;\; case\;\; of\;\; N \equiv 2 \pmod{4}\\
- \;\;\; for \;\; the\;\; case\;\; of\;\; N \equiv 0 \pmod{4}
\end{array}
\right.\;,
$$
and $\gamma$ is a self-dual cocycle which satisfies
$$
{}^{\#}(C\cap C)=1 ,\;\;\; {}^{\#}(any \cap C)=0,
$$
where, of course, $C$ is the Poincare dual of $\gamma$.
\newline Case 2) N : ood

Adapt the following assignment of the sign such that
$$
\Omega={\tilde W}_0 \alpha^0+{\tilde W}_1 \alpha^1+
\cdots {\tilde W}_{[N/2]-1} \alpha^{[N/2]-1}
\pm{\tilde W}_{[N/2]} \beta^{[N/2]}\cdots +{\tilde W}_{N-2} \beta^{N-2}
- {\tilde W}_{N-1} \beta^{N-1},
$$
where the sign of the middle term is
$$
\left\{
\begin{array}{c}
+ \;\;\; for \;\; the\;\; case\;\; of\;\; N \equiv 1 \pmod{4}\\
- \;\;\; for \;\; the\;\; case\;\; of\;\; N \equiv 3 \pmod{4}
\end{array}
\right.\;.
$$
}

\subsection{The (N-2)-point Yukawa coupling}
Once we have a concrete representation of the holomorphic (N-2,0)-form on M,
it is not so hard to give a concrete representation of the (N-2)-point
Yukawa coupling. Recall that the (N-2)-point Yukawa coupling is defined as
the following integral;
$$
\int_M \Omega \wedge \frac{\partial^{N-2}}{\partial \phi^{N-2}}\Omega \;.
$$
Substituting (\ref{Omg}) into (\ref{Y.C.}), we obtain
\begin{eqnarray}
\int_M \Omega \wedge \frac{\partial^{N-2}}{\partial \psi^{N-2}}\Omega
& =& \int_M \sum_i \tilde{W}^i \tilde{\alpha}^i \wedge
\frac{d^{N-2}}{d{\psi}^{N-2}} \sum_j \tilde{W}^j \tilde{\alpha}^j \nonumber \\
 & = & \sum_i \sum_j \tilde{W}^i \frac{d^{N-2}}{d{\psi}^{N-2}} \tilde{W}^j
\int_M \tilde{\alpha}^i \wedge \tilde{\alpha}^j \nonumber \\
& = & \sum_{i=0}^{N-2} (-1)^i \tilde{W}^i \frac{d^{N-2}}{d{\psi}^{N-2}}
\tilde{W}^{(N-2)-i}\;.
\end{eqnarray}
It would be a hard and boring calculation to carry out the above derivatives
and summations. But there is a smarter way of the {\it Wronskian}.
let us define the following formula ;
\begin{equation}
R_k = \sum_{i=0}^{N-2} \tilde{\omega}^i \frac{d^k}{d{\psi}^k}
\tilde{\omega}^{(N-2)-i}\;,
\label{defwrn}
\end{equation}
where $ \tilde{\omega}^i \equiv \frac{1}{\psi}\tilde{W}^i$.
(Recall that $\tilde{\omega}^i$'s satisfy the Picard-Fuchs equation (\ref{p4})
for the moduli variable $\psi$ and $\tilde{W}^i$'s satisfy the Picard-Fuchs
equation (\ref{p3}) for the moduli variable $z \equiv {\psi}^{-N}$.
The relation between the (N-2)-point Yukawa coupling defined by $W^i$ and
the one defined by ${\omega}^i$ is
\begin{eqnarray*}
& \sum_{i=0}^{N-2} (-1)^i \tilde{W}^i \frac{d^{N-2}}{d{\psi}^{N-2}}
\tilde{W}^{(N-2)-i} \\
= & \sum_{i=0}^{N-2} \psi \tilde{\omega}^i \frac{d^k}{d{\psi}^k}\psi
\tilde{\omega}^{(N-2)-i} \\
= & {\psi}^2 \sum_{i=0}^{N-2} \tilde{\omega}^i \frac{d^k}{d{\psi}^k}
\tilde{\omega}^{(N-2)-i}\;,
\end{eqnarray*}
here we have used Kodaira-Spencer theory .
The factor ${\psi}^2$ is not essential but only depends on the gauge choice
of $\Omega$. The correlation functions on the B-model take their values
on a section of the square of the line bundle on the moduli ${\cal M}$
in which the holomorphic (N-2,0)-form $\Omega$ lives \cite{candelas1}
\cite{candelas2}.)

$\tilde{\omega}^i$'s satisfy (\ref{p4}), that is,
$$ (1-{\psi}^N)\frac{d^{N-1}}{d {\psi}^{N-1}} \tilde{\omega}^i -
\sum_{l=1}^{N-2}b_l^{(N)}{\psi}^{l+1}\frac{d^l}{d {\psi}^l}\tilde{\omega}^i
= 0 .
$$
Multiplying l.h.s. by $\tilde{\omega}^{(N-2)-i}$ and summing up over
 i's, we obtain
\begin{equation}
(1-{\psi}^N)R_{N-1} -
\sum_{l=1}^{N-2}b_l^{(N)}{\psi}^{l+1}R_l
= 0.
\label{wronskian}
\end{equation}
Obviously from the definition (\ref{defwrn}), we can see that
$$ R_k = \int_M \Omega \wedge \frac{{\partial}^k}{\partial {\psi}^k} \Omega .
$$
 From the Kodaira-Spencer theory the top component of the
$\frac{{\partial}^k}{\partial {\psi}^k} \Omega$ is $H^{(N-2)-k,k} $.
Therefore
\begin{equation}
R_k=0\;\;\;for\;k<N-2\;.
\label{trv1}
\end{equation}
Thus nonzero terms in (\ref{wronskian}) are $R_{N-2}$ and $R_{N-1}$ alone.
In fact, there is a further trivial relation between $R_{N-2}$ and $R_{N-1}$
such that
\begin{equation}
R_{N-1}-\frac{(N-1)}{2}R^{\prime}_{N-2}=0\;,
\label{trv2}
\end{equation}
(where ${\prime}$ represents $d/d\psi$.)
For the time being we accept this.
Substituting (\ref{trv1}) and (\ref{trv2}) to (\ref{wronskian}),
we obtain a differrential equation of degree one such that
\begin{equation}
R^{\prime}_{N-2}-\frac{N{\psi}^{N-1}}{(1-{\psi}^N)}R_{N-2}=0\;.
\end{equation}
We can solve this very easily
$$
R_{N-2}=c\cdot \frac{1}{(1-{\psi}^N)}\;,
$$
where c is an integration constant. This is, of course, the very (N-2)-point
Yukawa coupling in the gauge of ${\omega}^i$'s. The (N-2)-point
Yukawa coupling in the gauge of $W^i$'s is
\begin{equation}
\sum_{i=0}^{N-2} (-1)^i \tilde{W}^i \frac{d^{N-2}}{d{\psi}^{N-2}}
\tilde{W}^{(N-2)-i} =
\frac{{\psi}^2}{(1-{\psi}^N)}.
\end{equation}
Here the integral constant is determined according to the concrete formula
of $\tilde{W}^i$'s (31).

It still remains to verify (\ref{trv2}). Recall that
$$ R_k = \sum (-1)^i \tilde{W}^i \frac{d^{k}}{d{\psi}^{k}}
\tilde{W}^{(N-2)-i}
$$
Then
\begin{eqnarray*}
R_k^{\prime} & = & \sum (-1)^i \tilde{W}^i \frac{d^{k+1}}{d{\psi}^{k+1}}
\tilde{W}^{(N-2)-i}+ \sum (-1)^i\frac{d}{d{\psi}} \tilde{W}^i
\frac{d^{k}}{d{\psi}^{k}} \tilde{W}^{(N-2)-i} \\
& \equiv & R_{k+1,0}+R_{k,1}\;.
\end{eqnarray*}
Here we have defined $R_{k,l}$ as
$$
R_{k,l}\equiv \sum (-1)^i\frac{d^l}{d{\psi}^l} \tilde{W}^i
\frac{d^{k}}{d{\psi}^{k}} \tilde{W}^{(N-2)-i}\;.
$$
Then we can obtain the following formula ;
\begin{eqnarray*}
R_k^{(l)} & = & R_{k+l,0}+l\cdot R_{K+l-1,1}+\cdots +R_{K,l} \\
          & = & \sum_{i=0}^l {}_lC_iR_{k+l-i,i}\;.
\end{eqnarray*}
We can verify by several trivial calculations that
\begin{eqnarray*}
\lefteqn {R_{N-1}-(N-1)R_{N-2}^{\prime}+\frac{(N-1)(N-2)}{2}
R_{N-2}^{\prime \prime }
-\cdots} \\
 &  & \equiv  \sum_{i=0}^{N-1}(-1)^i {}_{N-1}C_iR_{N-1-i}^{(i)} \\
 &  &  =     (-1)^{N-1}R_{0,N-1}
\end{eqnarray*}
For the case that N is odd , $(-1)^{N-1}=1$ and $ R_{0,N-1}=-R_{N-1,0}=R_{N-1}
$. For the case that N is even , $(-1)^{N-1}=-1$ and $ R_{0,N-1}=
R_{N-1,0}=R_{N-1}$. Noting that $R_{k}=0$ for $k\leq N-3$, we can conclude
in both cases that
$$
R_{N-1}=\frac{N-1}{2}R_{N-2}^{\prime}.
$$

\subsection{The mirror map and the translation into A-model}
Let us now construct a mirror map between the moduli space of
A and B models.
According to Morrison \cite{morrison} the mirror map can be obtained by
the following
process. Let the solution of the Picard-Fuchs equation which is regular at
maximally unipotent point, say, $z=0$ be $W^0$. And let $W^1$ be the solution
which has a singularity of $\log z$ at $z=0$. The mirrpr map is
\begin{equation}
t=\frac{W^1(\psi)}{W^0(\psi)}
\end{equation}
where $t$ is a coordinate of the moduli space of A-model on W being .
 In fact, it will turn out in the next section that this variable
$t$ coinside with the coupling constant of the A-model's Lagrangian (\ref{a1})

We will adapt his idea in the arbtrary dimensional case.

Now we want to translate the $(N-2)$-point Yukawa coupling of the B-model
to that on the A-model. The $(N-2)$-point Yukawa coupling of the B-model is
given by
\begin{equation}
\int_M \Omega \wedge \frac{\partial^{N-2}}{\partial \psi^{N-2}} \Omega
= \frac{\psi^2}{1-\psi^N}.
\end{equation}
We should note that the correlation function on the B-model is not a scalar on
 the moduli space but take the value on the square of the line bundle on which
the holomorphic $(N-2,0)$-form lives. Therfore we should consider not only the
effect of the transformation of the coordinate but also the gauge choice of
$\Omega$. Following Candelas et.al. and Morrison, we will adapt the gauge
;
$$
\Omega \to \frac{\Omega}{W^0}.
$$
Thus
\begin{eqnarray}
 & \int_M \Omega \wedge \frac{\partial^{N-2}}{\partial \psi^{N-2}} \Omega
= \frac{\psi^2}{1-\psi^N} \;\;\; of\;the\;B-model \nonumber \\
\to & \frac{1}{{(W^0)}^2}\frac{\psi^2}{1-\psi^N}{\frac{d\psi}{dt}}^{N-2}
\;\;\; of\;the\;A-model.
\label{q50}
\end{eqnarray}
Of cource an overall constant is not determined in (\ref{q50}).
We will fix this constant in the next section by looking at
{\it the large radius limit} of the theory.

\section{The A-model}
\subsection{lagrangian and weak coupling limit}

As we have explained in the previous section, Topological Sigma Model can be
obtained by twisting $N=2$ Supersymmetric Sigma Model on M. A-model
corresponds to A-twist,which turns $\psi_{+}^i$ and $\psi_{-}^{\bar{i}}$
in (\ref{sigmalag}) into $\chi^{i}$,$\chi^{\bar{i}}$, and
$\psi_{+}^{\bar{i}},\psi_{-}^i$
into $\psi_{z}^{\bar{i}},\psi_{\bar{z}}^{i}$. In other words, A-twist means
subtraction of half of $U(1)$ charge from world sheet spin quantum number.
Then we get the following Lagrangian for the A-model.

\begin{equation}
 L = 2t \int_{\Sigma} d^{2}z(\frac{1}{2}g_{IJ}\partial_{z}\phi^{I}
\partial_{\bar{z}}\phi^{J} + i\psi_{z}^{\bar{i}}D_{\bar{z}}\chi^{i}
g_{i\bar{i}} + i\psi_{\bar{z}}^{i}D_{z}\chi^{\bar{i}}g_{\bar{i}i}
-R_{i\bar{i}j\bar{j}}\psi_{\bar{z}}^{i}\psi_{z}^{\bar{i}}\chi^{j}\chi_{\bar{j}}
)
\label{a1}
\end{equation}

(\ref{a1}) is invariant under the infinitessimal
BRST - transformation,

\begin{eqnarray}
   & \delta\phi^{i} = i\alpha\chi^{i} \nonumber\\
   & \delta\phi^{\bar{i}} = i\alpha\chi^{\bar{i}} \nonumber\\
   & \delta \chi^{i} = \delta \chi^{\bar{i}} = 0 \nonumber\\
   & \delta\psi_{z}^{\bar{i}} = -\alpha\partial_{z}\phi^{\bar{i}}
         -i\alpha\chi^{\bar{j}}\Gamma^{\bar{i}}_{\bar{j}\bar{m}}
         \psi_{z}^{\bar{m}}
        \nonumber\\
   & \delta\psi_{\bar{z}}^{i} = -\alpha\partial_{\bar{z}}\phi^{i}
         -i\alpha\chi^{j}\Gamma^{i}_{jm}\psi_{\bar{z}}^{m}
\label{a2}
\end{eqnarray}

This invariance allows us to consider only BRST-invariant observables.
As in section2, we define BRST operator $Q$ by $\delta V =
 -i\alpha\{Q,V\}$ for any field V. Of course, $Q^2 = 0$.

 We can rewrite the lagrangian (\ref{a1}) modulo the $\psi$ equation of
motion as;

\begin{equation}
  L = it\int_{\Sigma}d^2z \{Q,V\} + t\int_{\Sigma}\Phi^{*}(e)
\label{a3}
\end{equation}

where

\begin{equation}
V = g_{i\bar{j}}(\psi_{z}^{\bar{i}}\partial_{\bar{z}}\phi^{j} + \partial_{z}
     \phi^{\bar{i}}\psi_{\bar{z}}^{j})
\label{a4}
\end{equation}

and

\begin{equation}
 \int_{\Sigma}\Phi^*(e) = \int_{\Sigma}(\partial_{z}\phi^{i}\partial_{\bar{z}}
     \phi^{\bar{j}}g_{i\bar{j}}- \partial_{\bar{z}}\phi^{i}
     \partial_{z}\phi^{\bar{j}}g_{i\bar{j}})
\label{a5}
\end{equation}
(\ref{a5}) is the integral of the pullback of the Kh\"aler form $e$ of M,
and it
depends only on the intersection number between $\Phi_{*}(\Sigma)$ and
$PD(e)$($PD(e)$ denotes the Poincare Dual of $e$),which
equals to the degree of $\Phi$. By an appropriate normalization
 of $g_{i\bar{j}}$,we have

\begin{equation}
\int_{\Sigma}\Phi^{*}(e) = n
\label{a6}
\end{equation}
where n is the degree of $\Phi$.

 Next, we consider the correlation function of BRST-invariant observables
$\{ {\cal O}_{i}\}$, i.e.
\begin{equation}
\langle \prod_{i=1}^{k} {\cal O}_{i}\rangle = \int {\cal D}\phi
 {\cal D}\psi{\cal D}\chi e^{-L} \prod_{i=1}^{k} {\cal O}_{i}
\label{a6}
\end{equation}
 We have seen $\int_{\Sigma}\Phi^{*}(e) = n $ and we decompose the
space of maps $\phi$ into different topological sectors
$\{B_{n}\}$ in each of which $deg(\Phi)$ is a fixed
integer.

 We can rewrite (\ref{a6}) as follows.
\begin{equation}
\langle \prod_{i=1}^{k} {\cal O}_{i}\rangle = \int {\cal D}\phi
 {\cal D}\psi{\cal D}\chi e^{-L} \prod_{i=1}^{k} {\cal O}_{i}
     = \sum_{n=0}^{\infty}e^{-nt}\int_{B_n} {\cal D}\phi
 {\cal D}\psi{\cal D}\chi e^{-it\int_{\Sigma}d^2 z \
    \{Q,V\}} \prod_{i=1}^{k} {\cal O}_{i}
\label{a7}
\end{equation}

 And we set

\begin{equation}
\langle \prod_{i=1}^{k} {\cal O}_{i}\rangle_{n}
   = \int_{B_n} {\cal D}\phi
 {\cal D}\psi{\cal D}\chi e^{-it\int_{\Sigma}d^2 z \
    \{Q,V\}} \prod_{i=1}^{k} {\cal O}_{i}
\label{a8}
\end{equation}
 We can easily see that $\int_{\Sigma}d^{2}z\{Q,V\} = \{Q,\int_{\Sigma}d^{2}
zV\}$,i.e. lagrangian is BRST exact. It follows from this and $\{Q,{\cal O}_i
\}=0$ that $\langle \prod_{i=1}^{k}{\cal O}_i\rangle_{n}$ doesn't depend on the
coupling constant t and we can take weak coupling limit
$t \rightarrow \infty $ in evaluating the path integral.

 In this limit,the saddle point approximation of the path integral
 becomes exact. Saddle points of the lagrangian are given by

\begin{equation}
 \partial_{\bar{z}}\phi^{i} = 0\;\;\; \partial_{z}\phi^{\bar{i}} = 0
\label{a9}
\end{equation}

 (\ref{a9}) shows that the path-integral is reduced to an integral over
 the moduli space
 of holomorphic maps from $\Sigma$ to $M$ of degree $n$. We denote
this space as ${\cal M}_{n}$. We will give more explicit form of
${\cal M}_{n}$
in a later section.

\subsection{The Ghost Number anomally and BRST observables }
 In the previous subsection, we come to the conclusion that path integral
(\ref{a6}) is reduced to an integral over ${\cal M}_{n}$ weighted by one loop
determinants of the non zero modes. But in general, there are fermion zero
modes  which are given as the solution of $D_{\bar{z}}\chi^{i}=
D_{z}\chi^{\bar{i}}=0$ and $D_{\bar{z}}\psi_{z}^{\bar{i}}=
D_{z}\psi_{\bar{z}}^{i} = 0$. Let $a_n$ (resp.$b_n$) be the number of $\chi$
(resp.$\psi$) zero modes. If $M$ is a Calabi-Yau manifold, we can see from
Riemann-Roch Theorem,

\begin{equation}
 w_n=a_n-b_n=2(1-g)
\label{a10}
\end{equation}
where $g$ is the genus of $\Sigma$.

 The existence of Fermion zero mode is understood as Ghost number anomally,
because lagrangian (\ref{a1}) classically conserves the ghost number.
  In path integration,these zero modes appear only in the integration measure
except in $\prod_{i=1}^k{\cal O}_i$,and the correlation function
$\langle \prod_{i=1}^k{\cal O}_i
\rangle_{n}$ vanishes unless the sum of
the ghost number of ${\cal O}_i$ is equal
to $w_n$.

 $D_{\bar{z}}\chi^{i}=0$ (resp.$D_{z}\chi^{\bar{i}}=0$) can be considered as
the linearization of the equation $\partial_{\bar{z}}\phi^{i}=0$
(resp.$\partial_{z}
\phi^{\bar{i}}=0$) and we can regard $\chi$ zero mode as $T{\cal M}_n$.

$w_n$ is usually called ``virtual dimension `` of ${\cal M}_n$.
 In generic case $b_n=0$ and $dim({\cal M}_n) = w_n$ holds. Then we have
$dim({\cal M}_n)=a_n$.


 BRST cohomology classes of the A-model are constructed from the de Rham
cohomology classes $H^{*}(M)$ of the manifold M.
 Let $W= W_{I_1I_2\cdots I_n}(\phi)d\phi^{I_1}\wedge\cdots\wedge d\phi^{I_n}$
 be an $n$ form on $M$. Then we define a corresponding local operator of the
A-Model,

\begin{equation}
{\cal O}_{W}(P) = W_{I_1\cdots I_n}\chi^{I_1}\cdots\chi^{I_n}(P)
\label{a11}
\end{equation}

 From (\ref{a2}) we have

\begin{equation}
\{Q,{\cal O}_{W}\}=-{\cal O}_{dW}
\label{a12}
\end{equation}

which shows that if $W \in H^{*}(M)$,${\cal O}_{W}(P)$ is BRST-closed.

\subsection{Evaluation of the Path Integral}
 Now we discuss how we can evaluate $\langle \prod_{i=1}^{k}{\cal O}_i \rangle
_n$ . We take ${\cal O}_i$ to be
${\cal O}_{W_i}$ which is induced from $W_i \in H^{*}(M)$. By adding
appropriate exact forms  we can make $W_i$ into the differential form
which has delta function support on $PD(W_i)$. Then
${\cal O}_{W_i}(P_i)$ is non zero only if

\begin{equation}
 \Phi(P_i) \in PD(W_i)
\label{a13}
\end{equation}

Then integration over ${\cal M}_{n}$ is restricted to
$\tilde{\cal M}_n$
, which consists of $\Phi \in {\cal M}_n$ satisfying (\ref{a13}). In
evaluating $\langle \prod_{i=1}^{k}{\cal O}_{W_i}\rangle_n$, (\ref{a13})
imposes $\sum_{i=1}^{k}dim(W_i)$ conditions, so $dim(\tilde{\cal M}_n) =
dim({\cal M}_n)- \sum_{i=1}^{k}dim(W_i) = w_n + b_n -\sum_{i=1}^{k}dim(W_i)$.
But from the fact that ghost number of ${\cal O}_{W_i}$ equals to $dim(W_i)$
(contribution from $\chi$) and anomally canselation condition,we have
$dim(\tilde{\cal M}_n) = b_n$. In generic case where $b_n=0$,
$\tilde{\cal M}_n$
turns into finite set of points. Then we perform an one loop integral
over each of these points. The result is a ratio of boson and fermion
determinants,which cancel each other. Then contributions to
$\langle \prod_{i=1}^{k} {\cal O}_{W_i} \rangle_{n}$ in the generic case equals
to the number of instantons which satisfies (\ref{a13}),i.e.

\begin{equation}
\langle \prod_{i=1}^{k} {\cal O}_{W_i} \rangle_{generic} =
\sharp \tilde{\cal M}_n
\label{a14}
\end{equation}

 When $dim(\tilde{\cal M}_n) = b_n \geq 1$,there are $b_n$ $\psi$ zero modes
which we can regard as the fiber of the vector bundle $\nu$ on $\tilde{\cal
M}_n$.
In this case, contributions to
$\langle \prod_{i=1}^{k} {\cal O}_{W_i} \rangle_{n}$are known as the
integration of Euler class $\chi(\nu)$ on $\tilde{\cal M}_n$. If we consider
$\nu$ as a 0-dimensional vector bundle on a point in the generic case, we can
apply the same logic there. We denote each component of $\tilde{\cal M}_n$ as
$\tilde{\cal M}_{n,m}$ and obtain

\begin{equation}
\langle \prod_{i=1}^{k} {\cal O}_{W_i} \rangle_{n} = \sum_{m}
\int_{\tilde{\cal M}_{n,m}}\chi(\nu)
\label{a15}
\end{equation}

Hence from (\ref{a7})

\begin{equation}
\langle \prod_{i=1}^{k} {\cal O}_{W_i} \rangle = \sum_{n=0}^{\infty}
\sum_{m=1}^{m_n}e^{-nt}\int_{\tilde{\cal M}_{n,m}}\chi(\nu)
\label{a16}
\end{equation}

 In algebraic geometry, generic instantons of degree $n$ corresponds to
irreducible maps of degree n, and instantons of degree n with non-zero $\psi$
 zero mode to reducible maps which are $j$-th multiple cover of irreducible
maps of degree $n/j$ $(j|n)$

 Let $\tilde{\cal M}_{n,j,m}$ be the $m$-th connected component of moduli
spaces which are $j$-th multiple cover of $n/j$-th irreducible instantons,
 and $\nu_{j,m}$ be vector bundle of $\psi$ zero modes on
$\tilde{\cal M}_{n,j,m}$. Then we have from (\ref{a16}),

\begin{equation}
\langle \prod_{i=1}^{k} {\cal O}_{W_i} \rangle = \sum_{n=0}^{\infty}
\sum_{j|n}\sum_{m=1}^{m_{n,j}}e^{-nt}\int_{\tilde{\cal M}_{n,j,m}}
\chi(\nu_{j,m})
\label{a17}
\end{equation}
\section{Inversion of Mirror Map and Calculation of N-2 point correlation
function of A-Model}

 Mirror symmetry at the level of correlation function level asserts that
with the mirror map
$t = t(\psi)$, correlation functions of A and B models coincide.
\begin{equation}
\langle \prod_{i=1}^{k}{\cal O}_{B_i}(\psi)\rangle_{B-Model}
= \langle \prod_{i=1}^{k}{\cal O}_{A_i}(t)\rangle_{A-Model}
\label{c1}
\end{equation}

${\cal O}_{A_i}$ is the mirror counterpart of ${\cal O}_{B_i}$. In our case,
${\cal O}_{B_i} = X_{1}X_{2}\cdots X_{N}$ which is only one element of
$H^{1}(\tilde{M_N},T\tilde{M_N})$ (marginal operator) and ${\cal O}_{A_i} =
{\cal O}_{e}$ where $e$ denotes K\"ahler form of $M_N$. Since $M_N$ is
embedded in $CP^{N-1}$,$e$ is the only one element of $H^{1,1}(M_N)$.

 From (43) and (\ref{c1}),we have

\begin{equation}
\langle {\cal O}_e {\cal O}_e \cdots {\cal O}_e (t)\rangle =
\frac{1}{W_0(\psi(t))^2}\frac{\psi^{2}(t)}{1-\psi^{N}(t)}
(\frac{d\psi}{dt})^{N-2}(const)
\label{c2}
\end{equation}

 Combining (31) and (41) we obtain

\begin{eqnarray}
t = N\log N-(\sum_{n=1}^{\infty} b_n z^n)/(\sum_{n=0}^{\infty} a_n z^n)
 - \log z \nonumber \\
 a_n = \frac{(Nn)!}{(n!)^{N}N^{Nn}},\;\;\;
 b_n = a_n(\sum_{i=1}^{n}\sum_{k=1}^{N-1}\frac{k}{i(Ni-k)})\nonumber\\
z = \frac{1}{\psi^{N}}
\label{c3}
\end{eqnarray}

 Substitution of $z^{-\frac{1}{N}}$ for $\psi$ in (\ref{c2}) leads to the
formula

\begin{equation}
 \langle {\cal O}_e {\cal O}_e \cdots {\cal O}_e \rangle =
 (const.) \frac{1}{W_0(z)^{2}}\frac{1}{1-z}(\frac{d}{dt}\log z)^{N-2}
\label{c4}
\end{equation}
 If we can represent $\log z$ as power series of $e^{-t}$,
$ \langle {\cal O}_e {\cal O}_e \cdots {\cal O}_e \rangle $ takes the form of
(\ref{a17}). This can be done as follows.
 First, we substitute $e^x$ for z in (\ref{c4}).

\begin{equation}
 -t = -N\log N + x + \sum_{n=1}^{\infty} c_n e^{nx}
\label{c5}
\end{equation}

where

\begin{equation}
\sum_{n=1}^{\infty}c_n z^{n} =
(\sum_{n=1}^{\infty}b_n z^{n})/(\sum_{n=0}^{\infty}a_n z^{n})
\end{equation}

 Then we assume the following expansion,

\begin{equation}
 x = -t + N\log N + \sum_{n=1}^{\infty}\gamma_{n}e^{-nt}
\label{c6}
\end{equation}
 $\gamma_{n}$ can be determined from compatibility of (\ref{c5})
 and (\ref{c6}).
 We put (\ref{c6}) into (\ref{c4}) and determine the constant with the
assumption that constant term of
$\langle {\cal O}_e {\cal O}_e \cdots {\cal O}_e (t) \rangle$ coincides the
classical value $\int e\wedge \cdots \wedge e = N$.

Then we have $\langle {\cal O}_e {\cal O}_e \cdots {\cal O}_e (t) \rangle$
represented in the form of (\ref{a17}),

\begin{eqnarray}
\lefteqn{\langle {\cal O}_e {\cal O}_e \cdots {\cal O}_e (t) \rangle}
 \nonumber\\
&& = N + n_{1}^{N}e^{-t} + n_{2}^{N}e^{-2t} + n_{3}^{N}e^{-3t} +\cdots
\nonumber\\
&& where \nonumber\\
&& n_{1}^{N} = N^{N+1}(1-2a_1 - {{N-2}\choose 2} (b_1))
          = N^{N+1}  - (N-2)N(N!)\sum_{i=1}^{N-1}\frac{N-i}{i} -2N(N!)
  \nonumber\\
&& n_{2}^{N} = N^{2N+1}\{1-2a_1-b_1 +3a_1^2 -2a_2 +2a_1b_1+ \nonumber\\
&& {{N-2}\choose 1}(-b_1+4a_1b_1+2b_1^2-2b_2) +
{{N-2}\choose 2} b_1^2\} \nonumber\\
&& n_{3}^{N} = N^{3N+1}\{1-2a_1-2b_1+3a_1^2+5a_1b_1+\frac{3}{2}b_1^2-2a_2-b_2
\nonumber\\
&& -4a_1^3-8a_1^2 b_1 -3a_1b_1^2 + 6a_1a_2 + 4a_2b_1 +2a_1b_2 -2a_3
\nonumber\\
&& + {{N-2}\choose 1}(-b_1+3b_1^2+4a_1b_1-2b_2-10a_1^2b_1-15a_1b_1^2
\nonumber\\
&& -\frac{9}{2}b_1^3+7a_1b_2+5a_2b_1+9b_1b_2-3b_3)\nonumber\\
&& + {{N-2}\choose 2}(b_1^2-6a_1b_1^2-4b_1^3+4b_1b_2)\nonumber\\
&& + {{N-2}\choose 3}(-b_1^3)\}
\label{c7}
\end{eqnarray}

 We see that in general $n^{N}_{k}$ has the structure
\begin{equation}
 n_{k}^{N} = N^{kN+1} - (correction\;\; terms)
\label{c8}
\end{equation}




\section{Geometrical interpretation }
 In this section, we will see how to interpret geometrically the
coefficients of the expansion
 of the (N-2) point correlation function.
\begin{equation}
 \langle\overbrace{{\cal O}_{e},\cdots,{\cal O}_{e}}^{\mbox{$N-2$
times}}\rangle
= N + n^{N}_1 \exp(- t) + n^{N}_2 \exp(-2t) + n^{N}_3 \exp(-3t) + \cdots
\label{j1}
\end{equation}
  Our main tool is the argument of  Witten which says
in the generic case that topological correlation function
counts the numbers of
holomolphic maps from
$CP^1$ to $M_N$ that satisfy following conditions.
\begin{equation}
 \phi :CP^1 \rightarrow M_N \;\;\;\;\phi(s_i,t_i) \in PD_i(e)
\label{j2}
\end{equation}
  $$ (i = 1,2,\cdots,N-2) \;  (s_i,t_i) \in CP^1 $$
 Subscripts i
means that $PD_i(e)$ and $PD_j(e)$ $(i \neq j)$ are homlogically equivalent
but different  as submanifolds.

 If $k \geq 2$, $n^N_k$ counts the number of irreducible maps of degree $k$
and $j$-th multiple covers of irreducible maps of degree $k/j$,
so we have to separate these two contributions. But in $k=1$ case,
all maps of degree 1 are irreducible and direct calculation of
$n^N_1$ from the definition (\ref{j2}) is easy.

 Now our strategy is the following:
\newline
1. We first consider an approximate method of computing $n^{N}_{k}$ and
derive its dominant term $N^{kN+1}$ (see(\ref{c8})).
\newline
2. We next perform an exact calculation of $n^{N}_{1}$ using the method of
algebraic geometry and Schubert calculus on $Gr(2,N)$. We reproduce precisely
its expression of (\ref{c7}).

 Since $M_N$ is embedded in $CP^{N-1}$, ${\cal M}^N_k$ is a submanifold of
$\bar{\cal M}^N_k$, moduli space of holomorphic maps
from $CP^1$ to $CP^{N-1}$. We can approximate it by a simple model as follows.

\begin{equation}
(s,t) \rightarrow ( \sum_{j=0}^k a^1_j s^j t^{k-j},
\sum_{j=0}^k a^2_j s^j t^{k-j},\ldots,\sum_{j=0}^k a^N_j s^j t^{k-j})
\label{j3}
\end{equation}
 Coordinates of $\bar {\cal M}^N_k$ are $(a^i_j)$, and its dimension is
$(k+1)N-1$ because we neglect the difference of over-all constant factor.
Then we can approximate it by $CP^{(k+1)N-1}$.

 $\phi \in \bar {\cal M}^N_k$ is an element of ${\cal M}^N_k$ if $\phi (s,t)
\in M_N$ for all $(s,t)$. This condition is translated into a set of
polynomial
constraints on $CP^{(k+1)N-1}$ as follows;
\begin{eqnarray}
      & \phi (s,t) \in M_N \; for \; all \; (s,t)  \nonumber \\
       \Longleftrightarrow \;&  \sum_{i=1}^N(\sum_{j=0}^k a^i_j s^j t^{k-j})^N
        = 0\; for \;all \;(s,t) \nonumber \\
       \Longleftrightarrow \;&  \sum_{m=0}^{kN} f^m(a^i_j)s^m t^{kN-m} = 0 \;
         for \;all \;(s,t)\nonumber \\
       \Longleftrightarrow \;&  f^m(a^i_j) = 0 \;\; (for \; m=0,1,\ldots,kN)
\label{j4}
\end{eqnarray}
where  $f^m(a^i_j)$ are homogeneous polynomials of degree N.

 Heuristically $f^m(a^i_j)$ are all independent, dimension of
${\cal M}^N_k$ is $(k+1)N-1-(kN+1)=N-2$, which is consistent with the result
of Riemann-Roch theorem.

 Next, we roughly evaluate $n^N_k$ using (\ref{j2}) and (\ref{j4}).

$PD_i(e)$ is the Poincare dual of the K\"ahler form $e$ of $M_N$ and it is
realized
as intersection of $M_N$ and hyperplane in $CP^{N-1}$.
$PD_i(e)$ and $PD_j(e)$ $(i\neq j)$ are homologically equivalent but
different submanifolds of $M_N$. Then we can set $PD_i(e)$ in $(\ref{j2})$
as follows;

\begin{equation}
          PD_i(e) = \{ (X_1,X_2,\ldots,X_N) | \sum_{k=1}^N X_k^N = 0 ,
            X_i = 0\}
\label{j5}
\end{equation}
               $$ (i=1,2,\ldots,N-2) $$

  and we set $(s_i,t_i)$'s in $CP^1$ in $(\ref{j2})$ as

\begin{eqnarray}
   (s_1,t_1) = (0,1), (s_2,t_2) = (1,0), (s_3,t_3) = (1,-1) \nonumber\\
and,\;\;(s_i,t_i) = (c_i      ,-1)\;\;  (i = 4,5,\ldots,N-2)
\label{j6}
\end{eqnarray}
        $$  (c_i \neq c_j  (i \neq j),\; and \; c_i \neq 1) $$

  Combining (\ref{j4}),(\ref{j5}),and (\ref{j6}), $n^N_k$ is roughly
calculated as the
number of points in $M^N_k$ which satisfy a system of algebraic equations,

\begin{equation}
     f^m(a^i_j) = 0 \; \; (m=0,1,2,\ldots,kN)
\label{j7}
\end{equation}

\begin{equation}
   \sum_{i=1}^N a^i_0 = 0\\, \sum_{i=1}^N a^i_k = 0,\\
 \sum_{i=1}^N \sum_{j=1}^k a^i_j (-1)^{k-j} = 0, \\
 \sum_{i=1}^N \sum_{j=1}^k a^i_j (c_m)^j (-1)^{k-j}
\label{j8}
\end{equation}

 We can't solve these equations explicitly, but we know that (\ref{j7})
and (\ref{j8}) are
homogeneous algebraic equations in $CP^{(k+1)N-1}$. So we naively apply
Bezout's theorem which says the number of solutions of homogeneous polynomial
 system in $CP^{(k+1)N-1}$ equals product of their homogeneous degree. In our
case, $f^m(a^i_j)$'s are degree $N$ and the rest are all degree $1$, and then
we have $n^N_k$ to be $N^{kN+1}$.

 Our approximate calculation differs from the formula (\ref{c7})
of the previous section. Indeed,
the above result only explains the leading term of $n^N_k$. This is because we
used a model too simple for $\bar {\cal M}^N_k$. One of the troubles
comes from
the fact that $CP^{(k+1)N-1}$ has many points which don't correspond to
holomorphic maps of degree $k$.

Example (1)
\begin{eqnarray*}
   CP^{N-1} \times CP^k & \to & CP^{(k+1)N-1}  \\
      (d_1,d_2,\ldots,d_N) \times (f_0,f_1,...,f_k)
  &  \mapsto &
\end{eqnarray*}
\begin{equation}
 (d_1(\sum_{j=0}^k f_j s^j t^{k-j}), d_2
(\sum_{j=0}^k f_j s^j t^{k-j}),\ldots,d_N(\sum_{j=0}^k f_j s^j t^{k-j}))  \\
  \simeq (d_1,d_2,\ldots ,d_N)
\label{j9}
\end{equation}

Example (2)
\begin{eqnarray*}
    CP^{2N-1} \times CP^{k-1} & \to & CP^{(k+1)N-1} \nonumber \\
      (d^1_0,d^1_1,d^2_0,d^2_1,\ldots,d^N_0,d^N_1) \times
(f_0,f_1,\ldots,f_{k-1}) & \mapsto &
\end{eqnarray*}
\begin{eqnarray}
((d^1_0 s + d^1_1 t)(\sum_{j=0}^{k-1} f_j s^j t^{k-1-j}),(d^2_0 s + d^2_1 t)
(\sum_{j=0}^{k-1} f_j s^j t^{k-1-j}),\ldots ,(d^N_0 s + d^N_1 t)
    (\sum_{j=0}^{k-1} f_j s^j t^{k-1-j})) \nonumber\\
                           \simeq  ((d^1_0 s + d^1_1 t),(d^2_0 s + d^2_1 t)
,\ldots,(d^N_0 s + d^N_1 t))
\label{j10}
\end{eqnarray}

 One can easily generalize the above maps.
\begin{eqnarray*}
   \eta_m : CP^{(m+1)N-1} \times CP^{k-m}& \to & CP^{(k+1)N-1} \nonumber \\
 (d^i_j)\;\;(i = 1,2,\ldots,N \;\; j= 0,1,\ldots,m) \times
(f_0,f_1,\ldots,f_{k-m})
& \mapsto &
\end{eqnarray*}
\begin{eqnarray}
((\sum_{j=0}^m d^1_j s^j t^{m-j})(\sum_{i=0}^{k-m} f_i s^i t^{k-m-i}),
\ldots,(\sum_{j=0}^m d^N_j s^j t^{m-j})(\sum_{i=0}^{k-m} f_i s^i t^{k-m-i}))
\nonumber\\
                          \simeq
((\sum_{j=0}^m d^1_j s^j t^{m-j}),(\sum_{j=0}^m d^2_j s^j t^{m-j}),
\ldots,(\sum_{j=0}^m d^N_j s^j t^{m-j}))
\label{j11}
\end{eqnarray}

 The image of the map $\eta_{i}$ corresponds to ,by projective equivalence, the
space of maps of degree $i$ in $CP^{(k+1)N-1}$,so we have to eliminate these
contribution from $N^{kN+1}$.Roughly speaking, that's one of the reason why
correction terms arises. But the situation is more complicated. Naive counting
of $dim(Im(\eta_i))$ concludes that it equals to
$((i+1)N-1)+(k-i)=(i+1)N+k-i-1$,while the condition for $\phi \in Im(\eta_i)$
to be a holomorphic map from $CP^1$ to $M_N$ only reduces the dimension of
$CP^{(i+1)N-1}$ from $((i+1)N-1)$ to $N-2$. This leads to the conclusion that
in $Im(\eta_i)$ $dim({\cal M}^N_k)$ is $N-2+k-i$. In other words, equations in
(\ref{j7}) are degenerate in $Im(\eta_i)$ and don't reduce the dimension
properly.

 One can easily see this situation if he tries to solve (\ref{j7}),(\ref{j8})
in the case of $N=3,k=1$.(There are no holomorphic maps of positive degree from
$CP^1$ to torus and $n^3_1=0$,so all the solutions of (\ref{j7}),(\ref{j8})
come from $Im(\eta_1)$.)

 Therefore, in order to construct ``exact''$\bar {\cal M}^N_k$ we have to
remove $Im(\eta_{k-1})$ from $CP^{(k+1)N-1}$.

 Degeneration of (\ref{j7}) also occurs when $\phi$ is the $j$-th multiple
cover of irreducible maps of degree $k/j$. In such a case,
 $\phi$ is decomposed
into the following form.

\begin{eqnarray}
\lefteqn{\phi = \tilde{\phi}\circ m_k} \nonumber\\
&& m_k : (s,t) \to (a_0 s^j + a_1 s^{j-1} t +\cdots + a_j t^j,
                     b_0 s^j + b_1 s^{j-1} t +\cdots + b_j t^j) \nonumber\\
&& \tilde{\phi} : (s,t) \to (a^{1}_{0} s^{\frac{k}{j}}+\cdots
                             +a^{1}_{\frac{k}{j}} t^{\frac{k}{j}},\ldots,
                             a^{N}_{0} s^{\frac{k}{j}}+\cdots
                             +a^{N}_{\frac{k}{j}} t^{\frac{k}{j}})
\label{aj7}
\end{eqnarray}
Then the condition $\phi (s,t) \in  M_N\;\; for\;\; all\;\; (s,t) \in CP^1$
reduces to
the one for $\tilde{\phi}$.

 Let ${\cal M}^{N}_{k,j}$(resp. $\bar{\cal M}^{N}_{k,j}$) be the set of maps
from $CP^1$ to $M_N$ (resp. from $CP^1$ to $CP^{N-1}$) which can be decomposed
into the form of (\ref{aj7}).$dim(\bar{\cal M}^{N}_{k,j})$ equals to
$(k/j+1)N-1+(2j-2)$.The above statement says , in
$\bar{\cal M}^{N}_{k,j}$, (\ref{j7}) acts only on the $\tilde{\phi}$
degrees of freedom. This leads us to the conclusion that
$ dim({\cal M}^{N}_{k,j}) = N-2 + (2j-2)$. The additional $2j-2$ degrees
of freedom which describes $j$-th multiple covers of $CP^{1}$ corresponds to
$2j-2 \;\;  \psi$ zero modes in the previous section. These are integrated and
leaves nontrivial contributions of order $n^{N}_{\frac{k}{j}}$ to $n^{N}_{k}$.
(The contribution from $\psi$ integration are considered to be order $1$.)

 These contributions are also the source of correction terms of $n^{N}_{k}$
and we have to treat them carefully in the case where $k \geq 2$.

 But in the $k=1$ case,troubles come only from the map $\eta_{0}$,and
 we eliminate it by using $Gr(2,N)$ instead of $CP^{2N-1}/Im(\eta_0)$.

 (An important difference between $Gr(2,N)$ and $(CP^{2N-1}/Im(\eta_0))$ is
that $Gr(2,N)$ is  the $SL(2,C)$ quotient space of the latter. Indeed
$dim(Gr(2,N))$ is $2N-4=(2N-1)-3$.)

 There is a map $\xi$
\begin{eqnarray}
  \xi : CP^{2N-1}/Im(\eta_0) & \to  Gr(2,N) \nonumber \\
         (a^1_0 s + a^1_1 t,\ldots,a^N_0 s + a^N_1 t)
                         & \mapsto
              \left(\begin{array}{cccc}
                    a^1_0& a^2_0& \ldots& a^N_0 \\
                    a^1_1& a^2_0& \ldots& a^N_0
                    \end{array}
                    \right) / GL(2,C)
\label{j12}
\end{eqnarray}
 Then, we have to decide the condition which corresponds to (\ref{j4}),i.e
the condition for $l \in Gr(2,N)$ to be contained in $M_(N)$.
 This condition can be translated into words of cohomolgy ring $H^{*}(Gr(2,N))$
{}.
 Let $F_N$ be the defining equation of $M_N$ and $s_N$ be the section
of $Sym^N(U^{*})$ defined from the restriction of $F_N$ to
$l \in Gr(2,N)$, where $U$ is the universal bundle of $Gr(2,N)$.

(See Appedinx A for the difinition of $s_N$)
 Then
\begin{eqnarray}
         &  l \in Gr(2,N)\;\; is\; contained\; in\; M_N \nonumber \\
                \Longleftrightarrow &  F_N|_l = 0 \nonumber \\
                \Longleftrightarrow  & s_N = 0 \; at\; l\in Gr(2,N) \nonumber
\\
                \Longleftrightarrow & l \in PD(c_T(Sym^N(U^*)))
\label{j13}
\end{eqnarray}

 In deriving last line from the third one, we used Gauss-Bonnet theorem
that says
zero locus of a section of vector bundle $E$ is homologically equivalent to
$PD(c_T(E))$.

 Since $rank(Sym^N(U^{*}))$ equals to $N+1$, $dim(PD(c_T(Sym^N(U^{*}))))$
 is $(2N-4)-N-1=N-5$, which agrees with $dim(\bar {\cal M}^N_k)=N-2$ and
$SL(2,C)$ equivalence.

 We can explicitly calculate $PD(c_T(Sym^N(U^{*})))$ in terms of Schubert
cycles of $Gr(2,N)$. Its construction depends on the fact that total Chern
class of $U$ is written as $c(U) = 1 - \sigma_1 t + \sigma_{1,1} t^2$ and
multiplication rules of $H^{*}(Gr(2,N))$ are well-known from two
theorems of Schubert cycles.
 (See  appndix B for construction.)

 After all, we have
\begin{eqnarray*}
\lefteqn{PD(c_T(Sym^N(U^{\star}))) =} \nonumber \\
& & N(N!) \sum_{i=0}^{\{N/2\}-2} Sym^{\{N/2\}-i}
(\beta_k) \sum_{j=0}^i ({2i\choose j}-{2i\choose j-1})\sigma_{\{N/2\}+i-j+1,
\{N/2\}-i+j+1} \nonumber \\ & & + N(N!)Sym^1(\beta_k) \sum_{j=1}^{\{N/2\}-1}
({N-3\choose j}-
{N-3\choose j-1})\sigma_{N-1-j,j+2} \nonumber \\ & &
+ N(N!) \sum_{j=2}^{\{N/2\}}({N-1\choose j}
-{N-1\choose j-1})\sigma_{N-j,j+1} \nonumber \\ & &
                                                            (for N: odd)
\end{eqnarray*}
\begin{eqnarray*}
\lefteqn{PD(c_T(Sym^N(U^{\star}))) =} \nonumber \\
& & N(N!) \sum_{i=0}^{\{N/2\}-2} Sym^{\{N/2\}-i}(\beta
_k) \sum_{j=0}^i ({2i+1\choose j}-{2i+1\choose j-1})\sigma_{\{N/2\}+i-j+1,\{N/2
\}-i+j+1} \nonumber \\ & &
+ N(N!)Sym^1(\beta_k) \sum_{j=1}^{\{N/2\}-1} ({N-3\choose j}-{N-3
\choose j-1})\sigma_{N-1-j,j+2} \nonumber\\ & &
+ N(N!) \sum_{j=2}^{\{N/2\}}({N-1\choose j}-
{N-1\choose j-1})\sigma_{N-j,j+1} \nonumber \\ & &
                                                            (for N: even)
\end{eqnarray*}
where
\begin{eqnarray}
                           \{N/2\} & = & N/2 - 1/2  (N: odd) \nonumber \\
                                   &   & N/2 - 1    (N: even)\nonumber \\
                           \beta_i & = & (N-2i)^2/((N-i)i)   \nonumber  \\
                    Sym^k(\beta_i) & := & \sum_{1\leq i_1<i_2<
                                     \cdots<i_k\leq\{N/2\}}
              {\beta_{i_1}}{\beta_{i_2}}\cdots{\beta_{i_k}}\\
                                   &    &  (i = 1,2,\ldots,\{N/2\}) .\nonumber
\label{j14}
\end{eqnarray}

 (\ref{j14}) represents the ``exact'' Moduli space of $\bar {\cal M}^N_1$
(devided by
$SL(2,C)$).

  Next, we have to decide the contributions to $n^N_1$ from a Schubert cycle
$\sigma_{N-j,j+1}$. We can calculate it combining (\ref{j7}) and (\ref{j12}).
By solving
(\ref{j7}) and using Map $\xi$,we can construct cycle $\alpha$ of dimension
$N+1$ in
 $Gr(2,N)$.
\begin{equation}
  \alpha := \left(\begin{array}{cccccccc}
               \gamma_1 & 0 & \gamma_3 & \gamma_4 & \ldots & \gamma_{N-2} &
\gamma_{N-1} & \gamma_N \\
                0 & \beta_2 & \gamma_3 & c_4\gamma_4 & \ldots &
c_{N-2}\gamma_{N-2} & \beta_{N-1} & \beta_N
              \end{array} \right)  
\label{j15}
\end{equation}
$$ (\beta_i,\gamma_i : arbitrary \; complex \;  number)  $$
 Contributions from $\sigma_{N-j,j+1} (j=2,3,\ldots,\{N/2\})$ to $n^N_1$ are
given by intersection number ${}^{\sharp}(\alpha \cap \sigma_{N-j,j+1})$. In
$j=2$ case we have a matrix representation of $\sigma_{N-2,3}$ as follows.
\begin{equation}
  \sigma_{N-2,3} = \left(\begin{array}{cccccccccc}
                    1 & 0 & 0 & 0 & \ldots & 0 & 0 & 0 & 0 & 0 \\
                    0 & \star & \star & \star & \ldots & \star & 1 & 0 & 0 & 0
              \end{array} \right)   
\label{j16}
\end{equation}
($\star$ represents arbitrary complex number and precisely speaking,
we have to add bondary
 points to (\ref{j16}) to compactify the cycle.)

 Then $(\alpha \cap \sigma_{N-j,j+1})$ are (in matrix representation
$dim(\alpha)+dim(\sigma_{N-2,3})=2N-4<2N=dim(Matrix)$,so we have to permit
multiplying each row of alpha by constant and adding one row to another when
we calculate intersection number)
\begin{eqnarray}
 \left(\begin{array}{ccccc}
          1 & 0 & 0 & \cdots & 0\\
          0 & 1 & 0 & \cdots & 0
          \end{array} \right)
        & ,\left(\begin{array}{cccccc}
            1 & 0 & 0 & 0 & \cdots & 0 \\
            0 & 0 & 1 & 0 & \cdots & 0
            \end{array} \right) \nonumber \\
               ,\ldots
           &  ,\left(\begin{array}{cccccccc}
                1 & 0 & \cdots & 0 & 0 & 0 & 0 & 0 \\
                0 & 0 & \cdots & 0 & 1 & 0 & 0 & 0
                \end{array} \right)  
\label{j17}
\end{eqnarray}
  Thus we get $ {}^{\sharp}(\alpha \cap \sigma_(N-2,3)) = N-5+1=N-4$.

We also have
\begin{equation}
      {}^{\sharp}(\alpha \cap \sigma_{N-j,j+1}) = N-2j    
\label{j18}
\end{equation}
   for general $j$ with some modification of the above method.

  (see Appendix C for detail)

 Combining (\ref{j14}) with (\ref{j18}),We get the following formula.

\begin{eqnarray}
\lefteqn{
{}^{\sharp}(PD(c_T(Sym^N(U^{*}))) \cap \alpha) =
}\nonumber \\
& &  N(N!) \sum_{i=0}^{\{N/2\}} Sym^{\{N/2\}-i}(\beta_k)
\sum_{j=0}^i ({2i\choose j}
-{2i\choose j-1})(2i-2j+1)
\nonumber \\
& &  - N(N!)Sym^1(\beta_k)(N-2)-N(N!)(N^2-3N+4)
\nonumber \\
& & (for N :odd)\nonumber \\
\lefteqn{
{}^{\sharp}(PD(c_T(Sym^N(U^{*}))) \cap \alpha) = }\nonumber \\
& &  N(N!) \sum_{i=0}^{\{N/2\}} Sym^{\{N/2\}-i}(\beta_k)
\sum_{j=0}^i ({2i+1\choose j}-{2i+1\choose j-1})(2i-2j+1)
\nonumber \\
& & -N(N!)Sym^1(\beta_k)
(N-2)-N(N!)(N^2-3N+4) \nonumber \\
& & {(for N :even)}
\label{j19}
\end{eqnarray}
 From an elementary identity of binomial coefficients
\begin{eqnarray}
   \sum_{j=0}^i ({2i\choose j} - {2i\choose j-1})(2j-2i+2) = 2^{2i}\nonumber \\
  \sum_{j=0}^i ({2i+1\choose j} - {2i+1\choose j-1})(2j-2i+2) = 2^{2i+1}
\label{j20}
\end{eqnarray}
 and
\begin{equation}
     \sum_{j=0}^{\{N/2\}} Sym^{\{N/2\}-j}(\beta_k/4) = \prod_{j=1}^
{\{N/2\}}
(1+ \beta_j/4) 
\label{j21}
\end{equation}
 We get
\begin{equation}
  {}^{\sharp}(PD(c_T(Sym^N(U^{\star}))) \cap \alpha) = N^{N+1} - N(N-2)(N!)
\sum_{j=1}^{N-1}((N-j)/j) - 2N(N!)    
\label{j22}
\end{equation}
   which agrees with the result from the calculation by the B-Model!
\section{Conclusion}

 We have done mainly two things in this paper. First,we generalize the results
of Candelas,de la Ossa,Greene,Parks and, Nagura,Sugiyama,which are the case of
$N = 3,4,5$.Recantly
Greene,Morrison and Plesser treated the same Model, and they calculated $3$
point function in addition to $N-2$ point function by analyzing monodromy of
Period Integral equation.   But our approach is slightly different from theirs.

 In our calculatoin with N left as parameter and we found certain structures
behind the expansion coefficients of $n^N_k$.

Secondly, we tried to explain the structure
 of $n^N_k$ by an elemantary algebro-geometrical approach, and in $k=1$ case we
 calculated $n^N_1$ exactly as the intersection number of cohomology of
Moduli space from the ``A-Model'' point of view. Our results coincides with
the results of the B-Model calculation, which leads us to the conclusion that
"Mirror Symmetry  " holds in general N-dimension.

 In $k \geq 2$ case, which we left untouched in this paper, we have two
approaches for geometrical
 interpretation of $n^N_k$. One is to calculate the contribution from
irreducible maps by generalizing the Grassmanian method and determine the
portion of multiple cover maps in $n^N_k$. The other is to use our
approximation method more accurately. We have derived only top term of
$n^N_k$ as yet, but we think some relation lies between the elimination of
$Im(\eta_j) (j \leq k)$ from $ CP^{(k+1)N-1}$ and the correction terms of
$n^N_k$.

\section*{Acknowledgement}
We sincerely thank Prof.~K.Ogiuiso {\it Department of Mathematics ,
Ochanomizu wemen's
University } and Dr.~K.Hori for helpful discussions especially on Shubert
calculus. We also thank Prof.~T.Eguchi for useful discussions and kind
encouragement.
\newpage
\section*{Appendix A}
     Construction of section $s_N$

 Let $ \sum_{j=1}^Nb_i^jX_j = 0 (i = 1,2,\ldots,N-2)$ be defining equations of
$l \in Gr(2,N)$. In matrix form , it can be writtten as
\begin{equation}
             BX = 0
\label{ja1}
\end{equation}
 where $B = (b_i^j)$ and $X = (X_j)$.

 We transform $B$ into a simple form by multiplying an $(N-2) \times (N-2)$
invertible matrix $D$ from the left and $N \times N$ invertible
matrix $C^{-1}$
from the right, i.e
\begin{equation}
      DBC^{-1} = \left(
                   \begin{array}{cccccccc}
                 0 & 0 & 1 & 0 & \cdots & \cdots & \cdots & 0\\
                 0 & 0 & 0 & 1 & 0 & \cdots & \cdots & 0\\
                 0 & 0 & 0 & 0 & 1 & 0 & \cdots & 0\\
                 \cdots & \cdots & \cdots & \cdots & \cdots & \cdots & \cdots &
\cdots \\
                 \cdots & \cdots & \cdots & \cdots & \cdots & \cdots & \cdots &
\cdots \\
                 0 & 0 & 0 & 0 & 0 & \cdots & 0 & 1
                 \end{array}
                 \right)
\label{ja2}
\end{equation}
 Then base transformation $CX=X'=(X'_j)$ turns (\ref{ja1}) into a form
\begin{equation}
             X'_i = 0 \;\;\; (i = 3,4,\ldots,N)
\label{ja3}
\end{equation}
 From (\ref{ja3}) we can see $X'_1$ and $X'_2$ as dual coordinate basis of
$l\in Gr(2,N)$.In other words, they are the basis of the fiber of $U^{*}$
at $l$.($U$ denotes universal bundle of $Gr(2,N)$.)

 Let $F_N = \sum_{i=1}^N X_i^N$ be defining equation of $M_N$. We substitute
$X_i$ in $F_N$ by $(C^{-1})_i^jX'_j$ and set $X'_i = 0 (i = 3,4,\ldots,N)$.
 (This operation corresponds to restriction of $F_N$ at $l$.)

 Then we get homogeneous polynomial of $X'_1$ and $X'_2$ of degree $N$, which
defines section of $Sym^N(U^{*})$ at $l$.

\section*{Appendix B}
\footnote{ We owe this part of discussion to Prof.Ogiuiso.}
Construction of $c_T(Sym^N(U^{\star}))$

{\large {Fact.}}
$c(U) = 1 - \sigma_{1}t + \sigma_{1,1}t^2$

 Hence we have $c(U^{*}) = 1 +
\sigma_{1}t + \sigma{1,1}t^2$, where $c(E)$ denotes total chern class of vector
 bundle $E$ and $U$ denotes universal bundle of $Gr(2,N)$.

 We formally represent $U^{*}$ as direct sum of line bundles $E$ and $F$
i.e. $U^{*} = E \oplus F$ and we set
\begin{eqnarray}       c(E) = 1 + xt \nonumber\\
                       c(F) = 1 + yt \nonumber\\
                       (x \;\;and\;\;y\; are\; formal\; variables.)
\label{jb2}
\end{eqnarray}
 From Fact and (\ref{jb2}),we have $c(U^{*}) = c(E)c(F) = 1 + (x+y)t +(xy)
t^2$,and
\begin{eqnarray}     x + y = \sigma_1 \nonumber\\
                        xy =  \sigma_{1,1}
\label{jb3}
\end{eqnarray}
 We can formally decompose $Sym^N(U^{*})$ into the form
\begin{equation}
    Sym^N(U^{*}) = E^{\otimes N}\oplus E^{\otimes N-1}\otimes F\oplus
\cdots
\oplus E^{\otimes N-1}\otimes F\oplus F^{\otimes N}
\label{jb4}
\end{equation}
 and we have
\begin{equation}
  c(Sym^N(U^{*})) = (1+Nxt)(1+((N-1)x+y)t)\cdots (1+Nyt)
\label{jb5}
\end{equation}
   Top Chern class is given as the coefficients of $t^{N+1}$.
\begin{equation}
  c_T(Sym^N(U^{*})) = Nx((N-1)x+y)((N-2)x+2y)\cdots Ny
\label{jb6}
\end{equation}
 $c_T(Sym^N(U^{*}))$ consists of symmetric polynnomials of $x$ and $y$,so
from $(B.3)$, we can represent $c_T(Sym^N(U^{*}))$ as polynomials of
$\sigma_1$ and $\sigma_{1,1}$.
 The result is
\begin{eqnarray}
    c_T(Sym^N(U^{*}))& = &N(N!)\sigma_{1,1}\prod_{j=1}^{\{N/2\}}\{
(\sigma_1)^2 + ((N-2j)^2/((N-j)j)\}\sigma_{1,1}\}\nonumber \\
& =& N(N!)\sum_{i=0}^{\{N/2\}} Sym(\{N/2\}-i)(\beta_k)\sigma_1^{2i}
\sigma_{1,1}^{\{N/2\}-i+1}\nonumber\\
       & &  (N:odd)\nonumber\\
      c_T(Sym^N(U^{*}))& =& N(N!)\sigma_1\sigma_{1,1}\prod_{j=1}^{\{N/2\}}\
{(\sigma_1)^2 + ((N-2j)^2/((N-j)j))\sigma_{1,1}\}}\nonumber\\
&=& N(N!)\sum{i=0}^{\{N/2\}}
Sym(\{N/2\}-i)(\beta_k)\sigma_1^{2i+1}\sigma_{1,1}^{\{N/2\}-i+1}\nonumber\\
  &  &   (N:even)
\label{jb7}
\end{eqnarray}

 Multiplication rules for Schubert cycles of $Gr(d,N)$ can be derived from two
fundamental theorems. See for proofs \cite{griffiths}.

 Pieri's  formula
\begin{eqnarray}
 If\; a = (a,0,0,\ldots,0),\; then\; for\; any\;
b = (b_1,b_2,b_3,\ldots,b_d) \nonumber\\
             (\sigma_a \cdot \sigma_b)
 =\sum_{(c_1,c_2,\ldots,c_d) :\sum c_j = \sum b_j + a \;  N-d \geq c_1 \geq b_1
\geq c_2 \geq \cdots \geq c_k \geq b_k}
\sigma_c
\label{jb8}
\end{eqnarray}

 Giambelli's formula
\begin{eqnarray}
 \sigma_{(a_1,a_2,a_3,...,a_d)} = det(s_{i,j}) \nonumber\\
   where\;\; s_{i,j} = \sigma_{a_i+j-i}
\label{jb9}
\end{eqnarray}
 From (\ref{jb8}) and (\ref{jb9}), we can derive two formula that are
true only for $H^{*}(Gr(2,N))$.
\begin{equation}
\sigma_{1,1}^n = \sigma_{n,n} \;\;\;     (n \leq N-2)
\label{jb10}
\end{equation}
\begin{eqnarray}
   \sigma_1^k \cdot \sigma_{n,n}& =& \sum_{i=0}^{[k/2]}({k\choose i} -
{k\choose i-1})\sigma_{k+n-i,n+i}   \;\;\;                     (k + n \leq N-2)
\nonumber\\
  \sigma_1^{N-1} \cdot \sigma_{1,1}& =& \sum_{i=2}^{[N/2]}({N-1\choose i} -
{N-1\choose i-1})\sigma_{N-i,i+1} \nonumber\\
  \sigma_1^{N-3} \cdot \sigma_{2,2}& =& \sum_{i=1}^{[N/2]-1}({N-3\choose i} -
{N-3\choose i-1}))\sigma_{N-1-i,i+2}
\label{jb11}
\end{eqnarray}

 Combination of (\ref{jb7}), (\ref{jb10}), and (\ref{jb11}) leads to the
formula
\begin{eqnarray}
\lefteqn{PD(c_T(Sym^N(U^{*}))) =}\nonumber \\
& & N(N!) \sum_{i=0}^{\{N/2\}-2} Sym^{\{N/2\}-i}(\beta_k) \sum_{j=0}^i
({2i\choose j}-{2i\choose j-1})\sigma_{\{N/2\}+i-j+1,\{N/2\}-i+j+1}\nonumber\\
& &  + N(N!)Sym^1(\beta_k) \sum_{j=1}^{\{N/2\}-1}
({N-3\choose j}-{N-3\choose j-1})\sigma_{N-1-j,j+2}\nonumber\\
& & + N(N!) \sum_{j=2}^{\{N/2\}}({N-1\choose j}-{N-1\choose j-1})
\sigma_{N-j,j+1}\nonumber
\\
& & (for N: odd) \nonumber\\
\lefteqn{PD(c_T(Sym^N(U^{*}))) =}\nonumber \\
& & N(N!) \sum_{i=0}^{\{N/2\}-2} Sym^{\{N/2\}-i}(\beta_k) \sum_{j=0}^i
({2i+1\choose j}-{2i+1\choose j-1})\sigma_{\{N/2\}+i-j+1,\{N/2\}-i+j+1}
\nonumber \\
& & + N(N!)Sym^1(\beta_k) \sum_{j=1}^{\{N/2\}-1}
 ({N-3\choose j}-{N-3\choose j-1})\sigma_{N-1-j,j+2}\nonumber\\
& & + N(N!) \sum_{j=2}^{\{N/2\}}({N-1\choose j}-{N-1\choose j-1})
\sigma_{N-j,j+1}
\nonumber\\
& &  (for N: even)
\label{jb12}
\end{eqnarray}

\section*{Appendix C}
\footnote{ We owe this part of discussion to Dr.~Hori}
 Counting of ${}^\sharp(c_T(PD(Sym^N(U^{*}))) \cap \alpha)$

 Since we have $c_T(Sym^N(U^{*}))$ written in terms of Schubert cycles, we
only have to determine the intersection number ${}^\sharp(\sigma_{N-i,i+1}
\cap \alpha)$.$(i = 2,3,\ldots,\{N/2\})$

 As we have said in the body of this paper, by solving $(\ref{j8})$ explicitly
and
using map $\xi$,we have $\alpha$ represented as $N+1$ dim subspace in the space
 of $2 \times N$ matrix.
\begin{equation}
  \alpha := \left(\begin{array}{cccccccc}
               \gamma_1 & 0 & \gamma_3 & \gamma_4 & \ldots & \gamma_{N-2} &
\gamma_{N-1} & \gamma_N \\
               0 & \beta_2 & \gamma_3 & c_4\gamma_4 & \ldots & c_{N-2}
\gamma_{N-2} & \beta_{N-1} & \beta_N
              \end{array} \right)  
\label{C.1}
\end{equation}
              $$  (\beta_i,\gamma_i : arbitrary\;\; cpx\;\; number)  $$
 On the other hand,$\sigma{N-i,i+1}$ can also be represented as $N-5$ dim
subspace as follows.
\begin{equation}
\left(\begin{array}{ccccccccccc}
               \star & \cdots & \star & 1 &
0 & 0 & 0 & 0 & 0 & \cdots & 0\\
             \star & \cdots & \star & 0 & \star & \cdots & \star & 1 & 0 &
\cdots & 0
        \end{array}         \right)
\label{C.2}
\end{equation}
(Precisely speaking, (\ref{C.2}) represents internal points subset of
$\sigma_{N-i,i+1}$, so we have to compactify it by adding boundary points.)

 Then, what we have to do is to determine the intersection points between
(\ref{C.1}) and (\ref{C.2}).

 We have two troubles:\newline
$1.$ (\ref{C.1}) and (\ref{C.2}) are in $2 \times N$ matrix form and $GL(2,C)$
indeterminate. So in counting intersection points,we can multiply each row
vector of (\ref{C.1}) by constant and add one row to the other.
\newline
$2.$ In $i \geq 3$ case,(\ref{C.1}) and (\ref{C.2}) do not intersect
transeversely,
i.e
intersect in more than one dimension, so we have to substitute $\alpha$ by the
cycles ${\alpha_i}$ $(i=2,3,\ldots,\{n/2\})$ which are homologically
equivalent to $\alpha$ and intersects transversely with $\sigma_{N-i,i+1}$.
$$
\alpha_i
=
\begin{array}{ccc}
\underbrace{
\left(
\begin{array}{cccc}
\alpha_3 & \alpha_4 & \cdots & \alpha_i \\
\alpha_3 & c_4 \alpha_3 & \cdots & c_i\alpha_i
\end{array}
\right.
}_{\mbox{{i-2}'s}}
 &
\begin{array}{cc}
\alpha_1 & 0 \\
0 & \beta_2
\end{array}
 &
\begin{array}{cccc}
\alpha_{i+1} & \cdots & \alpha_{N-i}+\alpha_i & \cdots
\\
c_{i+1}\alpha_{i+1} & \cdots & c_{N-i}\alpha_{N-i}+c_i\alpha_i & \cdots
\end{array}
\end{array}
$$
\begin{equation}
\left.
\begin{array}{cc}
\begin{array}{cc}
\alpha_3+\alpha_{N-3} & \alpha_{N-2}\\
\alpha_3+c_{N-3}\alpha_{N-3} & c_{N-2}\alpha_{N-2}
\end{array}
&
\begin{array}{cc}
\alpha_{N-1} & \alpha_N \\
c_{N-2}\alpha_{N-2} & c_{N-2}\alpha_{N-2}
\end{array}
\end{array}
\right)
\label{C.3}
\end{equation}


Then,$\sigma_{N-i,i+1}$ and $\alpha$ intersects in the following $N-2i$ points.
 $$\left(\begin{array}{cccccccc}
          0 & \cdots & 0 & 1 & 0 & \cdots & \cdots & 0\\
          0 & \cdots & 0 & 0 & 1 & 0 & \cdots & 0
          \end{array} \right)
         ,\left(\begin{array}{ccccccccc}
            0 & \cdots & 0 & 1 & 0 & 0 & \cdots & \cdots & 0 \\
            0 & \cdots & 0 & 0 & 0 & 1 & 0 & \cdots & O
            \end{array} \right)
$$
\begin{equation}
               ,\ldots
              ,\left(\begin{array}{ccccccccccc}
                0 & \cdots & 0 & 1 & 0 & \cdots & 0 & 0 & 0 & \cdots & 0 \\
            0 & \cdots & 0 & 0 & 0 & \cdots & 0 & 1 & 0 & \cdots & 0
                \end{array} \right)  
\label{C.4}
\end{equation}

(Notice that intersection points lie in boundary component of $\sigma_{N-i,i+1}
$ except for last one.)

 Finally we have ${}^{\sharp}(\sigma_{N-i,i+1} \cap \alpha) = N-2i$.


\newpage
\begin{thebibliography}{99}
\bibitem{nag} M.Nagura and K.Sugiyama, ``{\it Mirror Symmetry of K3 Surface}",
UT-663, to appear in Int.J. of Mod. Phys.
\bibitem{h.d.m.}B.R. Greene, D.R. Morrison and
M.R. Plesser, Mirror Manifolds in Higher Dimension, CLNS-93/1253,
IASSNS-HEP-94/2, YCTP-P31-92
\bibitem{w1} E.~Witten,
{\it Mirror Manifolds and Topological Field Theory}, in
{\it Essay on Mirror Manifolds}, ed. S.~-T.~Yau,
(Int. Press. Co., Hong Kong, 1992), pp.120-180.
\bibitem{ey}T.~Eguchi and S.T.Yang, Mod. Phys. Lett. A, Vol. 5, No. 21
(1990) 1693-1701.
\bibitem{ks1}K.Kodaira and D.C.Spencer, Ann. Math. 67 (1958) 382;
67 (1958) 403; 71 (1960) 43.
\bibitem{ks2}K.Kodaira, Complex manifold and Deformation of complex structure
( Iwanamishoten,{\it in japanese} 19?? or Springer,1985 )
\bibitem{l.s.w.} W.~Lerche, D.~Smit and N.~Warner,
Nucl.~Phys. {\bf B372} (1992) 87.
\bibitem{morrison}D.Morrison,``{\it Picards-Fuchs Equations and Mirror Maps
For Hypersursaces}", in ``{\it Essays on Mirror Manifolds}, ed. S.~-T.~Yau,
(Int. Press. Co., Hong Kong, 1992).
\bibitem{candelas1} P.~Candelas and X.~de la Ossa,
Nucl.~Phys. {\bf B355} (1991) 415.
\bibitem{candelas2} P.~Candelas, X.~de la Ossa, P.~Green and L.~Parkes,
Phys.~Lett. {\bf 258B} (1991) 118; Nucl.~Phys. {\bf B359} (1991) 21.
\bibitem{griffiths}P.~Griffiths and J.~Harrith, {\it Principles of Algebraic
Geometry}, ( Wiley, 1978 )
\bibitem{katz} S.~Katz ``{\it Rational curves on Calabi-Yau manifolds:
verifying predictions of Mirror Symmetry}'' Oklahoma State University
preprint OSU-M-92-3,1992
\end{thebibliography}
\end{document}